\documentclass[11pt,a4paper]{article}
\usepackage{jcappub}
\usepackage{color}

\title{Semi-empirical catalog of early-type galaxy-halo systems: dark matter density profiles, halo contraction and dark matter annihilation strength}

\author[a,b]{Kyu-Hyun Chae,}
\author[c,d]{Andrey V. Kravtsov,}
\author[b,c,d]{Joshua A. Frieman}
\author[e]{Mariangela Bernardi}
\affiliation[a]{Department of Astronomy and Space Science, Sejong University, 
 98 Gunja-dong Gwangjin-Gu, Seoul 143-747, Republic of Korea}
\affiliation[b]{Center for Particle Astrophysics, Fermi National Accelerator 
Laboratory, P.O. Box 500, Batavia, IL 60510, USA}
\affiliation[c]{Kavli Institute for Cosmological Physics, 5640 South Ellis 
Avenue, The University of Chicago, Chicago, IL 60637, USA}
\affiliation[d]{Department of Astronomy and Astrophysics, 5640 South Ellis 
Avenue, The University of Chicago, Chicago, IL 60637, USA}
\affiliation[e]{Department of Physics and Astronomy, University of Pennsylvania,
 209 South 33rd Street, Philadelphia, PA 19104, USA}

\emailAdd{chae@sejong.ac.kr}
\emailAdd{andrey@oddjob.uchicago.edu}
\emailAdd{frieman@fnal.gov}
\emailAdd{bernardm@physics.upenn.edu}

\abstract{
 With Sloan Digital Sky Survey galaxy data and halo data from
up-to-date N-body simulations within the
$\Lambda$CDM framework we construct a semi-empirical catalog (SEC) 
of early-type galaxy-halo systems by making a self-consistent bivariate 
statistical match of stellar mass ($M_{\star}$) and velocity dispersion 
($\sigma$) with halo virial mass ($M_{\rm vir}$) as demonstrated here for the 
first time. 
 We then assign stellar mass profile and velocity dispersion 
profile parameters to each system in the SEC using their observed correlations 
with $M_{\star}$ and  $\sigma$. Simultaneously, we solve for dark matter density
profile of each halo using the spherical Jeans equation. 
 The resulting dark matter density profiles deviate in general from the 
dissipationless profile of Navarro-Frenk-White or Einasto and their mean inner 
density slope and concentration vary systematically with $M_{\rm vir}$.
 Statistical tests of the distribution of profiles at fixed 
$M_{\rm vir}$ rule out the null hypothesis that it follows the 
distribution predicted by dissipationless N-body simulations for 
$M_{\rm vir} \lesssim 10^{13.5-14.5} {\rm M}_{\odot}$.
 These dark matter profiles imply that dark matter density is, 
on average, enhanced significantly in the inner region of halos with 
$M_{\rm vir} \lesssim 10^{13.5-14.5} {\rm M}_{\odot}$ supporting halo contraction.
 The main characteristics of halo contraction are: (1) the mean dark matter 
density within the effective radius has increased by a factor  
varying systematically up to $\approx 3-4$ at
$M_{\rm vir} = 10^{12} {\rm M}_{\odot}$, and 
(2)  the inner density slope has a mean of 
$\langle \alpha \rangle \approx 1.3$ with $\rho_{\rm dm}(r) \propto r^{-\alpha}$
and a halo-to-halo rms scatter of ${\rm rms}(\alpha) \sim 0.4-0.5$ 
for $10^{12}{\rm M}_{\odot}\lesssim M_{\rm vir}\lesssim 10^{13-14}{\rm M}_{\odot}$ 
steeper than the NFW profile ($\alpha=1$).
Based on our results we predict that halos of nearby elliptical and lenticular 
galaxies can, in principle, be promising targets for $\gamma$-ray
 emission from dark matter annihilation.}


\arxivnumber{1202.2716}

\begin{document}

\maketitle 

\flushbottom

\section{Introduction}

The currently standard cold dark matter (to be referred to as $\Lambda$CDM in 
conjunction with Einstein's cosmological constant $\Lambda$) model of structure
formation requires that galaxies form at the centers of dark matter 
halos \cite{WR,WF}. Dark matter halos embedding galaxies are indicated by
rotation curves of spiral galaxies \cite{Rub,Ken,Per}, velocity dispersions of
early-type (i.e.\ elliptical and lenticular) galaxies \cite{Ger,Dek,Cap}, and 
gravitational lensing effects induced by galaxies \cite{Rus,Koo,Bar} and galaxy 
clusters \cite{Clo,Kne}. In the $\Lambda$CDM model initial small density 
fluctuations grow by gravity and overdense regions 
 collapse to form halos dominated by CDM and halos 
merge subsequently to form larger and larger halos in a hierarchical fashion. 
These halos provide gravitational potential wells for baryonic matter to sink 
dissipationally to form visible galaxies in their centers. 
Current $\Lambda$CDM-based galaxy formation models reproduce well observed 
clustering properties of galaxy distributions and are ever improving towards
the goal of explaining all observed statistical and intrinsic properties of
galaxies \cite{Spr,Con,Zeh}.

Despite the fact that dark matter halos are pivotal for the standard 
($\Lambda$CDM) model of structure formation to explain observed galaxies,
our current knowledge of them is limited in two main aspects:
dark matter particles have never been identified conclusively
\cite{Bert,Fen,Por} and statistical properties of dark matter distribution 
within halos have not been characterized fully and robustly 
through observational studies \cite{Sch,Rey,Aug,Ogu}, 
hydrodynamic simulations \cite{Gne11,Duf,Abad,Tis,Joh} or empirical 
modeling of galaxy formation \cite{TG,Dut,BB}. 
In particular, detailed dark matter density profiles including both inner 
density slope and concentration have been obtained only for a limited
number of individual systems \cite{Xue,Sei,Son,Nap,Ume,New}.
Precise knowledge of the dark matter distribution within halos is crucial
for the search of dark matter particles because both direct and indirect 
detections of dark matter particles depend on the dark matter density 
within halos \cite{Por,McC}. 
The dark matter distribution within  halos is 
expected to be modified by dissipational galaxy formation in their centers from
the initial distribution predicted by dissipationless N-body simulations
\cite{Blu,Gne04,Gne11,Duf,Abad,Tis}. Hence, a precise statistical 
characterization of dark matter distribution is crucial for
galaxy formation physics as well \cite{Gne11,Duf}. Galaxy formation modeling
requires properly taking into account halo contraction effects. This means 
that observationally determined contracted halo mass profiles provide useful
constraints on galaxy formation physics including dissipational gas 
cooling/accretion, star formation/feedback, supernovae and AGN. 

Here we present a procedure of constructing a semi-empirical catalog of 
early-type galaxies and their embedding halos. We intend to construct a
realistic catalog that is useful for simulational studies such as lensing 
and at the same time produce a rigorous and general statistical 
characterization of halo mass profiles that is allowed by currently best 
statistical knowledge of galaxies and halos through a Jeans analysis. 
Our approach is to combine a statistically representative sample of observed
galaxies with halos from a cosmological dissipationless N-body simulation.
Specifically, we have two complete and separate sets of data for galaxies and 
halos:  photometric and spectroscopic data for galaxies \cite{Ber,Guo} 
 from the Sloan Digital Sky Survey (SDSS) \cite{Aba} 
and halos from the Bolshoi simulation \cite{Kly} (section~2). 
Based on the data sets we follow two steps 
to infer the dark matter distribution in halos. The first step is
to make a statistically consistent one-to-one match between galaxies and halos
for early-type systems by simultaneously assigning stellar masses ($M_{\star}$) 
and velocity dispersions ($\sigma$) to halo virial masses ($M_{\rm vir}$)
 so that observed statistical functions (of $M_{\star}$ and $\sigma$) and 
 correlations (between $M_{\star}$ and $\sigma$ and between $M_{\star}$ and 
$M_{\rm vir}$) are reproduced (section~3). Each galaxy with $M_{\star}$ and 
$\sigma$ assigned is further assigned  the effective radius ($R_{\rm{e}}$) and 
the S\'{e}rsic index ($n$) of the light distribution using their observed
 correlations with $M_{\star}$ and $\sigma$ (appendix~A). We refer
to the resulting catalog as a semi-empirical catalog (SEC) of early-type
galaxy-halo systems. 
The next step is to carry out Jeans dynamical modeling of each system with 
the assigned galaxy parameters and the unknown dark matter distribution 
(section~4). Based on the stellar mass profile and the SDSS 
velocity dispersion we can obtain only a degenerate set 
(i.e.\ a range) of dark matter density profiles 
for each system and we show some restricted results under special
assumptions (section~4.3.1). But, then using observational constraints on
  velocity dispersion profiles (VPs) we assign a VP to each galaxy and 
simultaneously a general-class dark matter density profile to the embedding halo
using the spherical Jeans equation so that the posterior
distribution of VPs matches the observed distribution (section~4.3.2).  
We test our results against physically and observationally
 well-motivated dynamical constraints (appendix~D).
We discuss the implications of our results for halo contraction (section~5) 
and dark matter annihilation strength (section~6). 
We conclude in section~7.

\section{Data: galaxies and dark halos}

According to the $\Lambda$CDM paradigm a galaxy is embedded near the center of
its host halo. The halo itself may be embedded in a larger halo or may embed
smaller halos referred to as subhalos. The larger host halo and the subhalos 
also embed galaxies at their centers. In this way one can identify a one-to-one
match between galaxies (including satellites) and halos (including subhalos)
\cite{Kra,Tas,VO,Con,Mos,Beh,Cha}. 
A reliable match requires unbiased samples of galaxies and halos. 

\subsection{Observed galaxies from the SDSS}

A large and unbiased galaxy sample is provided by the completed SDSS. 
This survey covers a large area of sky providing a low 
redshift galaxy sample unaffected by large scale structures  of the Universe 
\cite{Aba}. Galaxies are observed not only photometrically but also 
spectroscopically providing structural and kinematical parameters such as 
the total stellar mass $M_{\star}$, the stellar velocity dispersion $\sigma$, 
the effective radius $R_{\rm{e}}$, \cite{Ber} and the 
S\'{e}rsic index $n$ \cite{Guo}. The velocity dispersion $\sigma$ refers 
to the luminosity-weighted line-of-sight velocity dispersion (LOSVD) within 
$R_{\rm{e}}/8$ as described in \cite{Ber}.
Velocity dispersions are actually measured within a fixed aperture of radius 
$1.5$~arcsec. The fixed aperture corresponds to different physical scales that
are smaller than $R_{\rm e}$ for most galaxies. 
Measured velocity dispersions are then corrected
 to a fixed physical radius of $R_{\rm{e}}/8$ using a typically 
observed radial profile of luminosity-weighted LOSVDs within $R_{\rm{e}}$. 
This corrected velocity dispersion 
thus corresponds to the luminosity-weighted LOSVD within $R_{\rm{e}}/8$. 
  
Galaxies can be classified by eye inspection, color, spectral features, or
luminosity profile. None of these methods are perfect in separating 
galaxies by morphology. With the goal of dynamical modeling we select 
a galaxy population that consist mostly of early-type spheroidal 
systems which are predominantly velocity dispersion supported systems. 
Our selection is based on a luminosity profile 
concentration index $C_r$, which is the ratio of the radial scale which 
contains 90 percent of the Petrosian luminosity in the $r$-band to that 
which contains 50 percent. Spheroidal (early-type) galaxies are known to have 
more concentrated profiles compared with disk (late-type) galaxies. 
We use the criterion of $C_r > 2.86$ to select spheroidal galaxies \cite{Ber}.
Our working assumption is that for our selected galaxies any rotating disks, 
if present, can be ignored for the Jeans dynamical modeling. 

\subsection{Theoretical halos from the Bolshoi dark matter only simulation}

Cosmological N-body simulations can be used to produce samples of dark matter 
halos. The predicted statistics of these halos without galaxies are robust 
classical results of cosmological physics.
  The predicted statistical properties of halos, however, depend on
the adopted values of cosmological parameters. We use the Bolshoi simulation
\cite{Kly} that is based on the following up-to-date cosmological parameters 
of the flat $\Lambda$CDM cosmology:
$h=0.7$, $\Omega_{\rm m}=0.27$, $n_s=0.95$ and $\sigma_8 = 0.82$ consistent 
with Wilkinson Microwave Anisotropy Probe seven-year results \cite{Kom}. 
The Bolshoi simulation differs from the Millennium simulation \cite{Spr}
particularly in that the latter adopts a higher $\sigma_8 = 0.9$. 
The Bolshoi simulation uses $2048^3$ 
particles in a cosmological box of size $250$~$h^{-1}$~Mpc allowing a 
statistically representative complete catalog of halos with virial mass 
(defined below)
$1.5\times 10^{10}h^{-1}{\rm M}_{\odot} < M_{\rm vir} < 
 (1-2)\times 10^{15}h^{-1}{\rm M}_{\odot}$.

The Bolshoi simulation identifies both isolated halos (distinct halos) and
satellite halos (subhalos) embedded in larger halos and characterizes
their properties separately. For any mass range the halo population is 
dominated by distinct halos (from more than 70 percent to 100 percent) 
but we include subhalos for accuracy. 
The Bolshoi simulation provides the distribution of halos in 
maximum circular velocity.
The maximum circular velocity is found to be tightly correlated with the virial
 mass $M_{\rm vir}$ that is defined to be the mass bounded by the virial radius
within which the mean density is equal to the virial overdensity 
$\Delta_{\rm vir}$ times the mean cosmic density 
$\rho_{\rm m}=\Omega_{\rm m} \rho_{\rm crit}$ ($\rho_{\rm crit}$
being the critical density of the Universe). For the above flat $\Lambda$CDM 
cosmological model $\Delta_{\rm vir}\approx 360$  and  the 
virial radius $r_{\rm vir}$ is related to $M_{\rm vir}$ by
\begin{equation}
r_{\rm vir}= 206.9h^{-1}\left(\frac{M_{\rm vir}}{10^{12} h^{-1} M_{\odot}}\right)^{1/3}
 {\rm kpc}
\end{equation}
at redshift $z=0$ \cite{BN}. 
The halo mass function  then follows
from the combination of distinct halos and subhalos at fixed $M_{\rm vir}$.
The Bolshoi simulation does not sample halos 
beyond  $M_{\rm vir} \approx (1-2)\times 10^{15}h^{-1}{\rm M}_{\odot}$.
However, an analytic extrapolation of the halo mass function beyond the mass 
limit agrees with a larger volume simulation ( `MultiDark') results \cite{Pra}
based on the same cosmological parameters.

\section{Statistically matching galaxies with halos: a bivariate 
log-normal distribution of stellar mass and velocity dispersion 
as a function of halo mass for early-type systems}

Abundance matching of a galaxy parameter (usually luminosity or stellar mass) 
with halo mass has been widely used recently \cite{Kra,Tas,VO,Con,Mos,Beh,Cha}. 
For the early-type galaxy population we carry out a rigorous abundance matching
of two parameters (stellar mass and velocity dispersion) of a galaxy with its 
host halo mass. There will result a bivariate distribution of stellar mass and 
velocity dispersion as a function of halo virial mass.  
 The halo virial mass includes
all mass within the halo, i.e.\ $M_{\rm vir} = M_{\rm dm} + M_{\star}$ where
$M_{\rm dm}$ is the dark matter mass and $M_{\star}$ is the stellar mass of the
galaxy.\footnote{We ignore the mass of interstellar gas that is non-negligible
for spiral systems because our work is concerned only with spheroidal systems.
Black holes harboring at the galactic centers are also 
negligible for our analysis although we include them for completeness.}
The radial dark matter density profile within the halo with $M_{\rm dm}$ is 
the unknown that we want to solve for. 
The galaxy is characterized primarily by its stellar mass $M_{\star}$ and 
 velocity dispersion $\sigma$. The radial stellar mass density profile 
within the galaxy with $M_{\star}$ is modeled by a deprojected form of the 
S\'{e}rsic mass profile \cite{Ser}. The effective radius 
$R_{\rm e}$ and the S\'{e}rsic index $n$ can be assigned according to
the observed correlations with $M_{\star}$ and $\sigma$ (appendix~A).  
Thus, the primary task is to assign simultaneously $M_{\star}$ and $\sigma$ to 
$M_{\rm vir}$ for all halos in a statistically consistent manner.
 
The successful match between galaxies and halos (i.e.\ the assignment of
$M_{\star}$ and $\sigma$ to $M_{\rm vir}$) requires that the results satisfy all 
known statistical functions and correlations. These include the halo mass
function, the galaxy stellar mass function, the galaxy velocity dispersion
function and the observed correlations between $M_{\rm vir}$ and $M_{\star}$ and 
between $M_{\star}$ and $\sigma$.\footnote{Note that no direct observational
constraint is available for the correlation between $M_{\rm vir}$ and $\sigma$.}
 The simultaneous assignment of $M_{\star}$ and $\sigma$ to $M_{\rm vir}$ 
can be done through a bivariate probability distribution 
of $M_{\star}$ and $\sigma$ as a function of 
$M_{\rm vir}$. The observed distribution of $\sigma$ 
at fixed $M_{\star}$ (or vice versa) is well described by 
a log-normal distribution for the early-type galaxy population \cite{Ber,HB}.  
The observed distribution of $M_{\star}$ at fixed $M_{\rm vir}$ can also be
described by a log-normal distribution \cite{Mor}.
These observations justify a choice of a bivariate normal distribution for the
logarithmic values of $M_{\star}$ and $\sigma$ at fixed $M_{\rm vir}$. Note that
a bivariate normal distribution is not an accurate model for the total galaxy
population, which includes both spheroidal and disk galaxies, because the 
observed distribution of $\sigma$ at fixed $M_{\star}$
for all galaxies is not log-normal but asymmetric. 

We derive the bivariate normal distribution for the logarithmic values of 
$M_{\star}$ and $\sigma$ as a function of $M_{\rm vir}$ for the early-type galaxy 
population using an iterative procedure described in detail below. 
The parameters of the bivariate
normal distribution are the mean $M_{\rm vir}$-$M_{\star}$ relation, 
 the mean $M_{\rm vir}$-$\sigma$ relation, the standard deviations of 
$\log_{10}(M_{\star})$ and  $\log_{10}(\sigma)$, and the correlation
coefficient between $\log_{10}(M_{\star})$ and  $\log_{10}(\sigma)$
as functions of $M_{\rm vir}$. The mean relations are based on an 
abundance-matching relation between $M_{\rm vir}$ and $M_{\star}$ and the SDSS 
observed relation between $M_{\star}$ and $\sigma$. Starting with initial 
guesses of the standard deviations and the correlation coefficient we iterate
keeping the mean relations fixed until the bivariate distribution reproduces
 the SDSS galaxy stellar mass and velocity dispersion functions and the 
$M_{\star}$-$\sigma$ relation up to the observational uncertainties.

\subsection{The detailed procedure}

We describe in detail the procedure\footnote{The reader who is only 
interested in the result may skip this lengthy subsection.} 
of generating a bivariate log-normal
distribution of stellar mass ($M_{\star}$) and stellar velocity dispersion
($\sigma$) as a function of halo virial mass ($M_{\rm vir}$), or
 a bivariate normal distribution of $Y\equiv\log_{10}(M_{\star}/M_{\odot})$ and 
$Z\equiv\log_{10}(\sigma/{\rm{km}}~{\rm{s}}^{-1})$ as a 
function of $X\equiv\log_{10}(M_{\rm vir}/M_{\odot})$. 
 We use a sufficiently large comoving volume of $4\times 10^9$~Mpc$^3$ 
to keep high precision in statistical match toward the large mass limit.
Our results are then limited only by the accuracies of the input
statistical quantities.  Throughout a system refers to the combination of 
a halo (whether it is isolated/distinct or a subhalo embedded in a larger halo) 
and the central galaxy embedded at the center of the halo.  

The statistical quantities of our use for galaxies  are 
(1) the stellar mass function (SMF) for all-type galaxies, (2) the 
 SMF and the velocity dispersion function (VDF) for early-type galaxies, (3) 
the number fraction of early-type galaxies $f_E$ as a function of $M_{\star}$,
and (4) the distribution of early-type galaxies in the stellar mass-velocity
dispersion plane, to be referred to as the $M_{\star}$-$\sigma$ relation.
The $M_{\star}$-$\sigma$ relation for early-type galaxies with $C_r > 2.86$ is 
derived and presented here while all the other have been published \cite{Ber}. 
Stellar masses are based on the Chabrier initial mass function throughout.
The statistical quantity of our use for halos is the halo mass function (HMF) 
that includes both distinct halos and subhalos. Throughout a statistical 
function (i.e.\ SMF, VDF, or HMF) is denoted by $\phi$ and expressed as
comoving number density per unit logarithmic interval of the variable under 
consideration. 

\begin{figure}
\begin{center}
\setlength{\unitlength}{1cm}
\begin{picture}(15,8)(0,0)
\put(-0.7,10.){\includegraphics{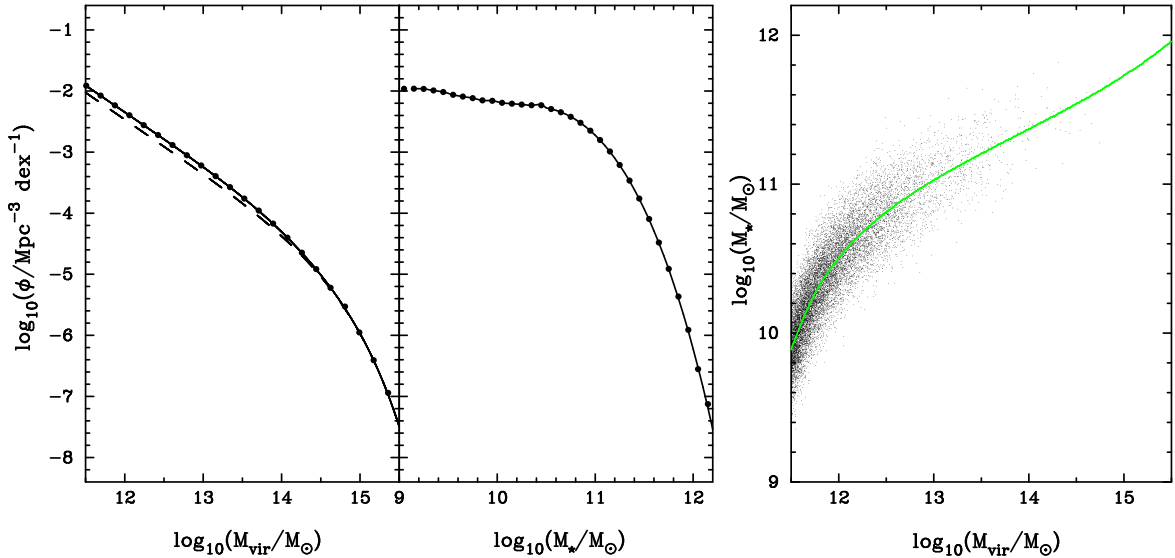}}
\end{picture} 
\caption{The first panel shows the halo mass function from the Bolshoi
N-body simulation. The solid curve includes all halos (i.e.\ both distinct and 
subhalos) while the dashed curve 
includes only distinct halos. Data points show mock halos
within a comoving volume of $4\times 10^9$~Mpc$^3$. The second panel shows
the SDSS galaxy stellar mass function for all-type galaxies. The curve is
the SDSS measured function while data points are from galaxies assigned to
the halos according to the relation shown in the last panel. 
The last panel shows the final abundance matching $M_{\rm vir}$-$M_\star$
 relation with a Gaussian intrinsic (constant) scatter of $0.17$ for 
$\log(M_\star)$ at fixed $M_{\rm vir}$.
}
\label{MvirMs}
\end{center}
\end{figure}

1. The first step is to find a mean relation between $M_{\star}$ and $M_{\rm vir}$
for all-type galaxies from the SMF and the HMF using the
abundance matching method \cite{Kra,Tas,VO,Con,Mos,Beh,Cha}. 
This method uses a prior knowledge (or assumption) of the intrinsic
scatter of one variable at the other. Observational studies find 
$s_Y=0.16$-$0.17$ for the intrinsic scatter of $Y$ at fixed $X$ insensitive to
the value of $X$ \cite{Mor,Mor2,Yan}.
 We adopt $s_Y=0.17$  \cite{Mor}.
We follow the procedure demonstrated in \cite{Cha}. Initially we obtain
an approximate abundance matching relation assuming zero intrinsic scatter.
This initial relation is biased at large $X$ owing to the ignored scatter.
The difference between the initial relation and the unknown true relation is
to be referred to as bias. The remaining task is to estimate the bias and thus
 the true mean relation as well.
We do this iteratively using a Monte-Carlo method in the following way. 
Let us first consider the initial mean relation as a surrogate of the true
mean relation. We generate halos from the HMF and then assign $Y$ to $X$ 
through the surrogate mean $X$-$Y$ relation and the intrinsic scatter 
$s_Y=0.17$ assuming the Gaussian model. From these mock systems we derive
a mock SMF which is of course quite different from the observed SMF 
because of the biased input surrogate relation. 
Now obtain a mock abundance matching relation between this mock SMF
and the HMF ignoring the intrinsic scatter. Then the difference between this
mock relation and the input surrogate relation is our first estimate of the 
bias. Using this bias we correct the initial abundance matching relation 
between the observe SMF and the HMF. With the corrected relation we do the
simulation all over again. There results a better corrected relation. 
We iterate this simulation until the mock SMF matches the observed SMF.
Usually a few iterations suffice.
Fig.~\ref{MvirMs} shows the reproduced HMF and SMF and the final 
$X$-$Y$ plane. The total number of the generated systems with 
$M_{\rm vir} \ge 10^{10.5}~{\rm M}_{\odot}$ is $\simeq 1.866 \times 10^8$.

\begin{figure}
\begin{center}
\setlength{\unitlength}{1cm}
\begin{picture}(11,14)(0,0)
\put(-1.3,-3.){\includegraphics{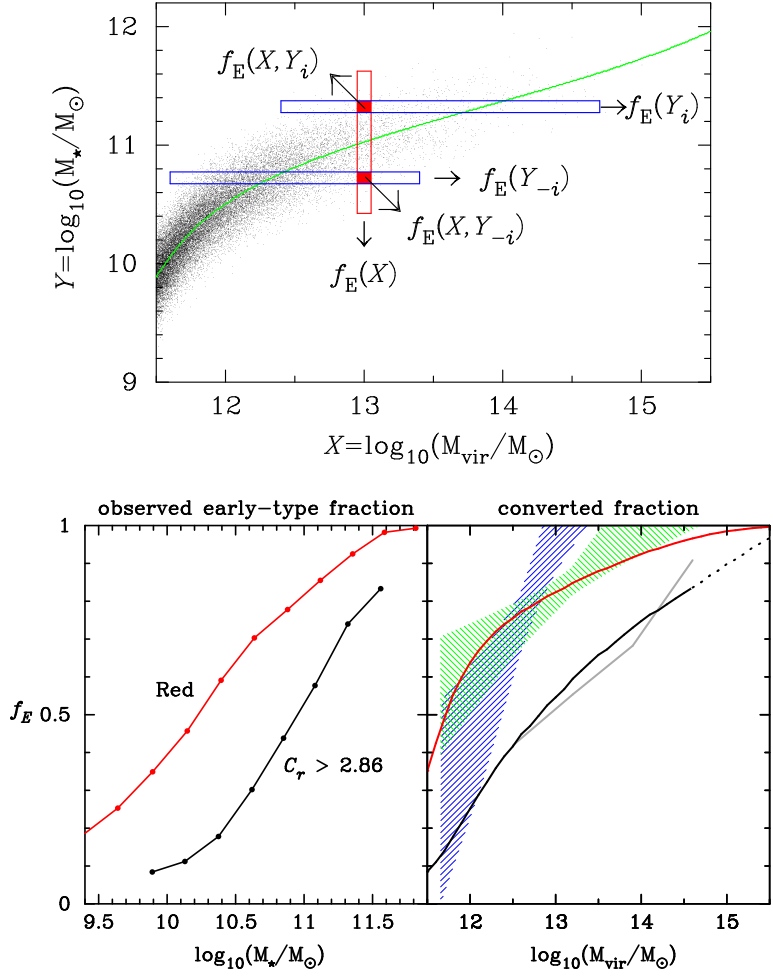}}
\end{picture} 
\caption{
(Top) A schematic view of converting the early-type fraction as a function of
 $Y$ to that at fixed $X$. We use 
$f_{\rm E}(X)=\sum_i [f_{\rm E}(X,Y_i) w_i + f_{\rm E}(X,Y_{-i}) w_{-i}]
\approx \sum_i[f_{\rm E}(Y_i)+f_{\rm E}(Y_{-i})] w_i$ where $w_i$ and
 $w_{-i}$ are statistical weights satisfying $\sum_i( w_i + w_{-i}) =1$. 
The last approximation follows from $w_{-i}=w_{i}$ and the assumption that 
$\{[f_{\rm E}(X,Y_i)-f_{\rm E}(Y_i)]+[f_{\rm E}(X,Y_{-i})-f_{\rm E}(Y_{-i})]\}w_{i}
\approx 0$, which is plausible (see the texts). 
(Bottom) The left panel shows
 the early-type number fraction (black curve) of SDSS galaxies as a function of
stellar mass. We use the observed fraction of galaxies with $C_r > 2.86$ 
for $10^{9.9} M_\odot \le M_{\star} \le 10^{11.6}M_\odot$  since outside this range 
the measured fraction is less reliable \cite{Ber}. The red galaxy number
fraction \cite{Ber} is also shown for comparison.
The right panel shows
 the early-type number fraction (black solid curve) as a function of 
halo mass ($M_{\rm vir}$) converted from the observed fraction at $M_{\star}$
using the $M_{\rm vir}$-$M_{\star}$ plane shown above. 
The dotted curve is a linear extrapolation as a 
function of $\log_{10} (M_{\rm vir})$ for the high-mass part 
($M_{\rm vir}>10^{14.5}M_\odot$).
The gray curve is the number fraction of visually selected early-type
galaxies based on halo occupation statistics results \cite{Guo}. 
The red curve shows the converted red number fraction for
our red galaxies. 
The hatched regions show observational results (95\% confidence regions) 
on the red fraction based on SDSS data and satellite kinematics 
(green/blue region based on the observed 
fraction as a function of luminosity/stellar mass) \cite{Mor}.
There are consistencies for both early-type fractions and red fractions.
}
\label{fE}
\end{center}
\end{figure}

2. We use the above $X$-$Y$ plane to convert the observed early-type 
fraction of systems as a function of $Y$ (stellar mass) to that at fixed $X$ 
(halo mass). Let $f_{\rm E}(X,Y)$ be the early-type fraction as a function of
$X$ and $Y$.  If $f_{\rm E}(X,Y)$ were known, $f_{\rm E}(X)$, 
the fraction at $X$ would be simply $\int f_{\rm E}(X,Y)P(Y|X) dY$ where 
$P(Y|X)$ is the probability distribution of $Y$ at $X$ that comes from the
$X$-$Y$ relation. In our case $P(Y|X)$ is the Gaussian distribution with 
a standard deviation of $s_Y = 0.17$.  Without $f_{\rm E}(X,Y)$ we use the 
following approximation 
\begin{equation}
f_{\rm E}(X) \approx \int f_{\rm E}(Y)P(Y|X) dY,  
\end{equation}
where $f_{\rm E}(Y)$ is the observed fraction at $Y$.
This approximation is obtained as follows:
\begin{eqnarray}
f_{\rm E}(X) & =  & \int f_{\rm E}(X,Y) P(Y|X) dY  \nonumber \\
        & \rightarrow & \sum_i [f_{\rm E}(X,Y_i) P(Y_i | X)
                   + f_{\rm E}(X,Y_{-i}) P(Y_{-i} | X)] \Delta Y \nonumber \\
    &=& \sum_i [f_{\rm E}(X,Y_i) w_i + f_{\rm E}(X,Y_{-i}) w_{-i}],  
\end{eqnarray} 
where $Y_i$ ($i=1,2,3,\cdots$) are greater than the mean $\langle Y \rangle$, 
 $Y_{-i}=2 \langle Y \rangle -Y_i$ (see the top panel of Fig.~\ref{fE}), and 
statistical weights 
$w_i [\equiv P(Y_i|X) \Delta Y]$ and 
$w_{-i} [\equiv P(Y_{-i}|X) \Delta Y]$ satisfy $\sum_i(w_i + w_{-i})=1$ and 
$w_{-i}=w_i$ as well for the symmetric Gaussian distribution.
It then follows
\begin{eqnarray}
f_{\rm E}(X) &=&  \sum_i [f_{\rm E}(X,Y_i) + f_{\rm E}(X,Y_{-i})] w_i \nonumber \\
       & = &  \sum_i [f_{\rm E}(Y_i) + f_{\rm E}(Y_{-i})] w_i 
          + \sum_i [\delta f_{\rm E}(X,Y_i) + \delta f_{\rm E}(X,Y_{-i})] w_i,
\end{eqnarray}  
where $f_{\rm E}(Y) = \int f(X,Y)P(X|Y)dX$ [$P(X|Y)$ being the probability
distribution of $X$ at fixed $Y$] and
 $\delta f_{\rm E}(X,Y) \equiv f_{\rm E}(X,Y) - f_{\rm E}(Y)$. But, we expect
$\delta f_{\rm E}(X,Y_i) \approx -\delta f_{\rm E}(X,Y_{-i})$ for $Y_i$ close to
 $\langle Y \rangle$ (i.e.\ small values of $i$) assuming that
$f_{\rm E}(X,Y)$ is a smooth distribution and the shape of $P(X|Y)$ is slowly 
varying with $Y$. 
For large values of $i$ the approximation
$\delta f_{\rm E}(X,Y_i)\approx-\delta f_{\rm E}(X,Y_{-i})$ may not be good enough
but $w_i \approx 0$ for the Gaussian distribution of $Y$. Hence, we have
$\sum_i [\delta f_{\rm E}(X,Y_i) + \delta f_{\rm E}(X,Y_{-i})] w_i \approx 0$ and
equation~(3.3) is reduced to equation~(3.1).
 
\begin{figure}
\begin{center}
\setlength{\unitlength}{1cm}
\begin{picture}(8,10)(0,0)
\put(-2.5,-3.){\includegraphics{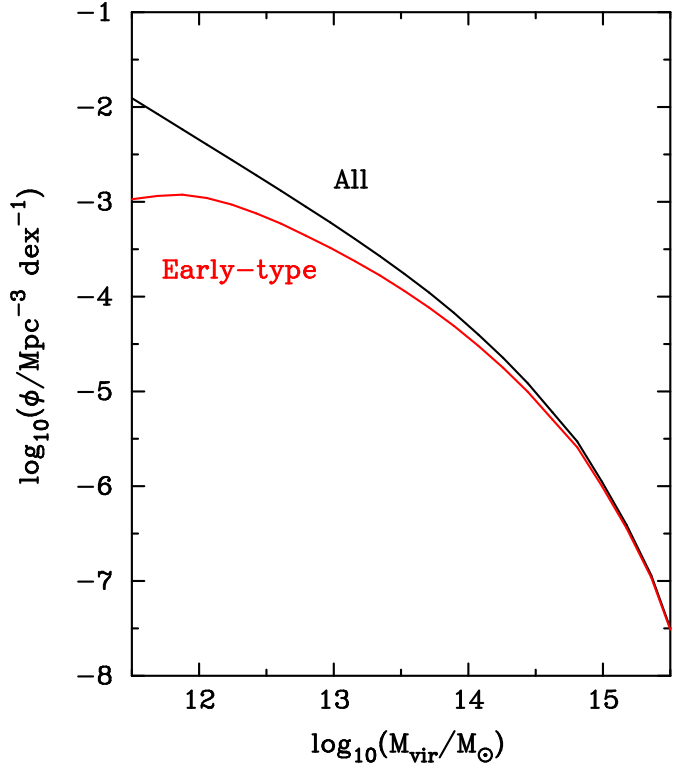}}
\end{picture} 
\caption{
Halo mass function for those halos embedding early-type galaxies at 
their centers (red curve) derived by multiplying the total halo mass function
(black curve) by the early-type fraction (shown in Fig.~\ref{fE}) as a function
 of $M_{\rm vir}$.
}
\label{HMFE}
\end{center}
\end{figure}

 Fig.~\ref{fE} shows the functional behaviors of observed and converted 
fractions. The early-type ($C_r>2.86$) fraction is compared with the red 
fraction. At a given stellar mass the red fraction is significantly higher than
 the early-type fraction. This implies that significant fractions of red 
galaxies do not possess concentrated light distributions.
 The converted fractions as functions of $M_{\rm vir}$ are compared with
recent independent results. Notice that our early-type fraction based on
luminosity concentration index is in good agreement with a fraction based on 
visual inspection of galaxy morphology \cite{Guo}. 
Our red fraction is also consistent
with independent results based on color selection \cite{Mor}. 
We obtain the HMF for early-type systems by multiplying the
number density for all-type halos by $f_{\rm E}(X)$.  Fig.~\ref{HMFE} shows 
the HMF for early-type systems.

\begin{figure}
\begin{center}
\setlength{\unitlength}{1cm}
\begin{picture}(15,8)(0,0)
\put(-.7,10.){\includegraphics{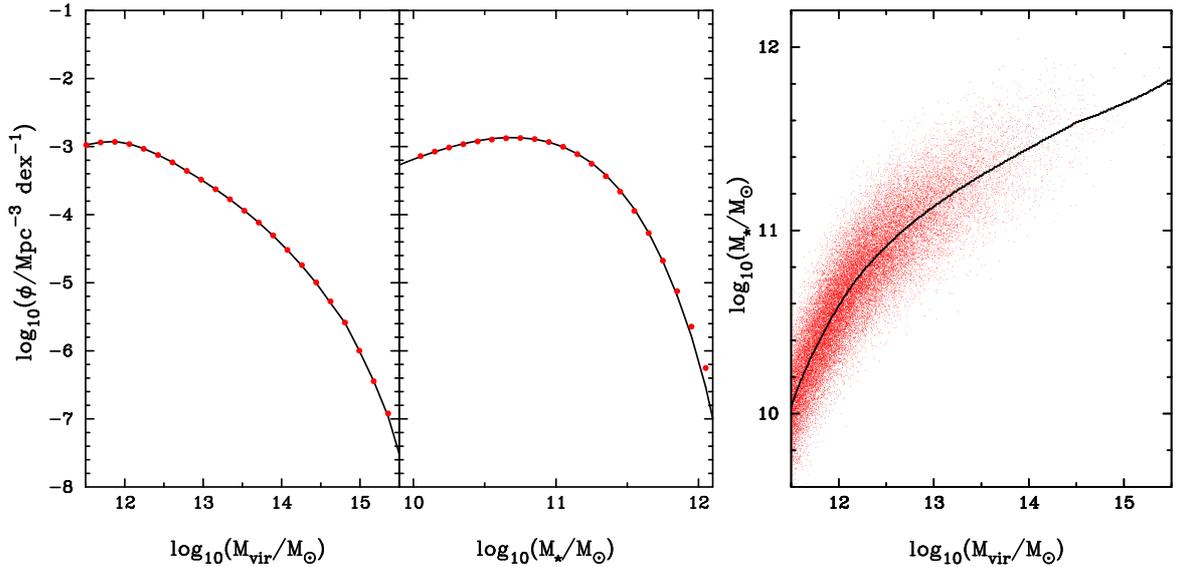}}
\end{picture} 
\caption{Same as Fig.~\ref{MvirMs} but for early-type galaxy-halo systems.}
\label{MvirMsE}
\end{center}
\end{figure}

3. Assign $Y$ to $X$ for the early-type population using the same procedure of
step~1 based on the early-type HMF from step~2 and the observed early-type SMF.
Fig.~\ref{MvirMsE} shows the resulting HMF and SMF and the $X$-$Y$ plane for the
 early-type population. We obtain $\simeq 6.9\times 10^6$ early-type systems 
with $X$ and $Y$ assigned.

\begin{figure}
\begin{center}
\setlength{\unitlength}{1cm}
\begin{picture}(15,8)(0,0)
\put(-0.7,10.){\includegraphics{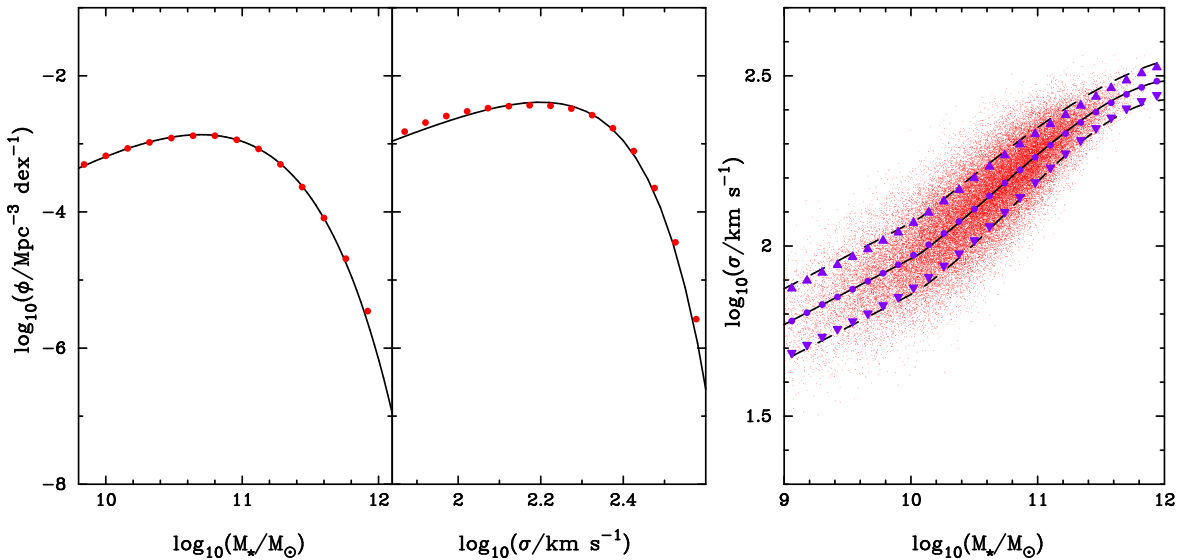}}
\end{picture} 
\caption{(Left)  The black curves are the SDSS stellar mass ($M_\star$) 
and stellar velocity dispersion ($\sigma$) functions for early-type galaxies
 with $C_r > 2.86$ \cite{Ber}. Red points are derived from the data points 
shown in the right panel. (Right) The black full and dashed curves represent 
our measured median values and standard deviations for SDSS early-type galaxies 
based on the data analyzed in \cite{Ber}. Red data points show a sample of 
mock galaxies produced from our final bivariate
distribution of $M_\star$ and $\sigma$ as a function of $M_{\rm vir}$ to be shown
 in section~3.2. The purple filled circles and upright/upside-down
triangles are respectively the median values and standard deviations of the
data points.
}
\label{MsVD}
\end{center}
\end{figure}

4. For the early-type systems from step~3, we assign $Z$ to $Y$ using only the 
observed $Y$-$Z$ relation (Fig.~\ref{MsVD}) ignoring the correlation between 
$X$ and $Z$. In this way we obtain an approximate set $\{X,Y,Z\}$. 
This set is approximate in the sense that parameter correlations and 
scatters have not yet been determined.
Nevertheless the mean relations among three parameters should be accurate 
as long as the distributions in the two-parameter planes are assumed symmetric
[observations support the symmetric distributions of $Y$ (at fixed $X$) and $Z$
(at fixed $Y$) for the early-type population]. This set allows us to 
obtain a crude estimate of $s_Z$ (standard deviation of $Z$) at fixed $X$.

5. Let us now assume that $Y$ and $Z$ follow a bivariate normal distribution 
at fixed $X$. The probability density function is given by
\begin{equation}
P(Y,Z|X)= \frac{1}{2\pi s_Y s_Z\sqrt{1-\rho_{YZ}^2}} 
 \exp\left[-\frac{1}{2(1-\rho_{YZ}^2)}g(Y,Z)\right] 
\end{equation}
with
\begin{equation}
g(Y,Z)=\frac{(Y-\mu_Y)^2}{s_Y^2}+\frac{(Z-\mu_Z)^2}{s_Z^2}
      -\frac{2\rho_{YZ}(Y-\mu_Y)(Z-\mu_Z)}{s_Y s_Z},
\end{equation}
where $\mu_Y$ and $\mu_Z$ are the mean values and $\rho_{YZ}$ is the correlation
coefficient at fixed $X$. The values of $\mu_Y$ and $\mu_Z$ come from step~4.
We assume $s_Y = 0.17$ \cite{Mor} independent of $X$. Parameters $s_Z$ 
and $\rho_{YZ}$ are the unknowns to be determined. With the first guess of 
$s_Z$ from step~4 and an initial guess of $\rho_{YZ}$ (e.g.\ $0.4$)
 we generate our first mock set including the relevant correlation. 
This mock set reproduces the SMF because we are using the
bias corrected abundance matching $X$-$Y$ relation. However,
the set does not reproduce
the observed VDF and the observed stellar mass-velocity dispersion
relation because of the inaccurate $s_Z$ and the unknown $\rho_{YZ}$.

6. We use the mock set itself to estimate the bias in $s_Z$. This is done
in the following way. First, calculate the standard deviation of $Z$ at fixed
$Y$, $s_Z(Y)$. Next, take the lower dimensional set $\{X,Y\}$ from the mock
set and then assign $Z$ to $Y$ using  $s_Z(Y)$ just calculated, pretending
the correlation in the mock set were unknown. Thus, we get a biased set
$\{X,Y,Z\}$. Finally, we calculate $s_Z$ at fixed $X$ from this biased set
and then compare it with that of the mock set to estimate the bias. 

7. Correct $s_Z$ for the bias estimated in step~6. Using this corrected
$s_Z$ generate a revised mock set $\{X,Y,Z\}$ using the bivariate normal
probability distribution of $Y$ and $Z$ at fixed $X$. We can check that this
revised mock set results in an improved match to the observed VDF. We note that
the resulting VDF has much to do with $s_Z$ but little to do with 
$\rho_{YZ}$. The correlation $\rho_{YZ}$ has much to do with the resulting 
$Y$-$Z$ relation, its mean relation and the dispersion of $Z$ at fixed $Y$.
Hence, for the given $s_Z(X)$ we adjust $\rho_{YZ}$ so that the resulting 
$Y$-$Z$ relation matches the observed stellar mass-velocity dispersion as 
closely as possible. For the sake of simplicity we allow $\rho_{YZ}$ to be only
a linear function of $X$ with broken slopes.

\begin{figure}
\begin{center}
\setlength{\unitlength}{1cm}
\begin{picture}(15,11)(0,0)
\put(-0.6,12.){\includegraphics{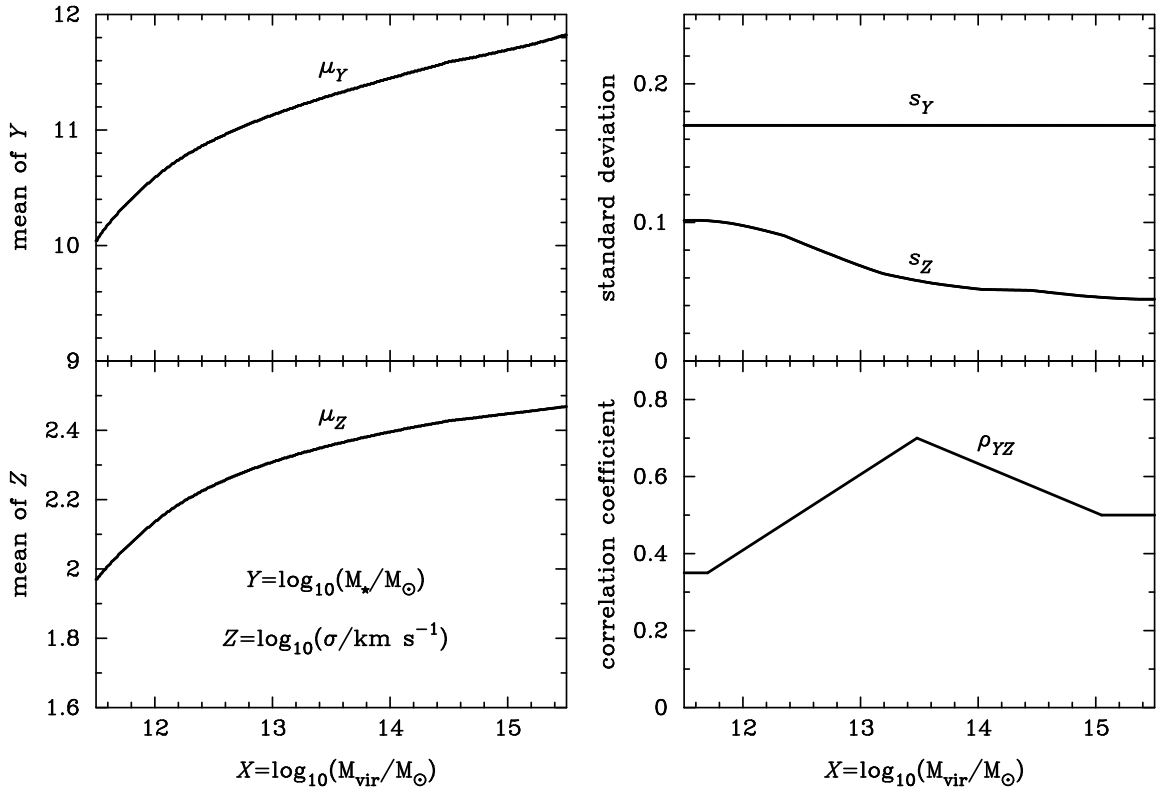}}
\end{picture}
\caption{
Parameters of the bivariate normal distribution of 
$Y$~[$\equiv \log_{10}(M_\star/M_\odot)$] and 
$Z$~[$\equiv \log_{10}(\sigma/{\rm km}~{\rm s}^{-1})$] as functions of
$X$~[$\equiv \log_{10}(M_{\rm vir}/M_\odot)$].
}
\label{bivpara}
\end{center}
\end{figure}

8. If the results of step~7 are not satisfactory go back to step~6 and iterate.
Initially we focus on reproducing the VDF by correcting $s_Z(X)$ with a
constant $\rho_{YZ}$. Eventually the initially adopted $\rho_{YZ}$ will be 
modified as the observed median relation between stellar mass and velocity 
dispersion cannot be well reproduced with a constant $\rho_{YZ}$. 
We find that with a constant $\rho_{YZ}=0.35$ the resulting VDF matches well 
the observed VDF for $\sigma > 100~{\rm km}~{\rm s}^{-1}$. This means that
through the iteration we have found a proper functional behavior of 
$s_Z(X)$ (see Fig.~\ref{bivpara}). On the other hand, the resulting mean
value of $Z$ at $Y$ deviates downward from the observed relation towards the 
high mass end. This requires us to correct the constant $\rho_{YZ}$. After
trial and error we find that a varying $\rho_{YZ}(X)$ with enhanced values
for relatively more massive systems, as shown in Fig.~\ref{bivpara}, can 
reproduce the observed median relation between stellar mass and velocity 
dispersion (see Fig.~\ref{MsVD}). However, for this model of $\rho_{YZ}(X)$ the
 intrinsic scatter of $Z$ at fixed $Y$ is slightly underestimated as can be 
seen in Fig.~\ref{MsVD}.
This lowered dispersion is a consequence of the strengthened correlation.
This manifests the difficulty of matching the VDF and the
stellar mass-velocity dispersion relation simultaneously to a high precision.
Some compromise is necessary and we have chosen to give more weight to the
VDF for $\sigma > 100~{\rm km}~{\rm s}^{-1}$ and the median 
stellar mass-velocity relation but less weight to the the intrinsic scatter
of velocity dispersion at fixed stellar mass as the observational error of
the latter has not been quantified.

\subsection{The result: a semi-empirical catalog of early-type galaxy-halo systems with some parameters undetermined}

\begin{figure}
\begin{center}
\setlength{\unitlength}{1cm}
\begin{picture}(15,15)(0,0)
\put(-1.,-5.){\includegraphics{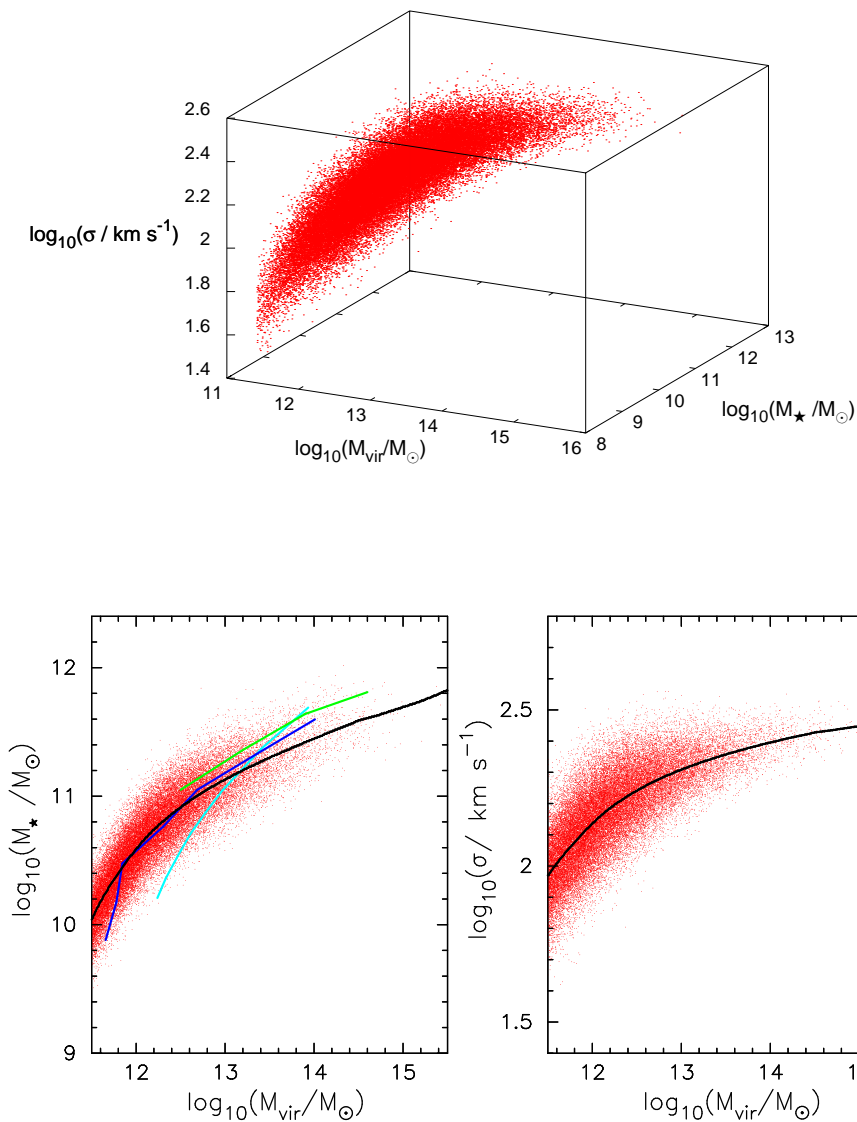}}
\end{picture}
\caption{(Top) The bivariate normal distribution of $\log_{10}(M_\star/M_\odot)$ 
and $\log_{10}(\sigma/{\rm km}~{\rm s}^{-1})$ as a function of  
$\log_{10}(M_{\rm vir}/M_\odot)$ for early-type galaxy-halo systems where
$M_{\rm vir}$ is the virial mass of the halo embedding an early-type galaxy
at the center and $M_\star$ and $\sigma$ are the stellar mass and central 
stellar velocity dispersion of the galaxy respectively.
(Bottom) Two projections of the three-dimensional distribution. Black curves are
the mean relations. The mean $M_{\rm vir}$-$M_\star$ relation is compared with
recent observational results on early-type or red galaxies based on 
galaxy-galaxy weak lensing (blue curve) \cite{Man}, 
halo occupation statistics (green) \cite{Guo} 
and satellite kinematics (cyan) \cite{Mor}.
}
\label{biv}
\end{center}
\end{figure}

 We obtain a final set \{$M_{\rm vir}$,~$M_{\star}$,~$\sigma$\} of 
$\simeq 6.9 \times 10^6$ mock early-type systems with
 $M_{\rm vir}>10^{10.5} M_{\odot}$ for a training comoving volume of 
$4 \times 10^9$~Mpc$^3$. A subset of $50,000$ systems is displayed in 
Fig.~\ref{biv}.
The projected $M_{\rm vir}$-$M_{\star}$ relation is compared with recent available 
observational results based on other methods including galaxy-galaxy weak 
lensing  \cite{Man}, halo occupation statistics  \cite{Guo}, 
and kinematics of satellite galaxies  \cite{Mor}. 
Our mean $M_{\rm vir}$-$M_{\star}$ relation lies 
near the median of other results for $M_{\rm vir}<10^{13} M_{\odot}$
but lies lower down to $-0.2$~dex for $M_{\rm vir}>10^{13} M_{\odot}$.
We note that this high mass part behavior of the abundance matching 
$M_{\rm vir}$-$M_{\star}$ relation in comparison to other methods is also found 
for all-type galaxies \cite{Beh}. However,
it turns out that this difference has relatively minor effect on our results 
on dark matter density profiles. The $M_{\rm vir}$-$\sigma$ relation for 
early-type galaxies is derived here for the first time. 
Abundance matching $M_{\rm vir}$-$\sigma$ 
relations for all-type galaxies can be found in \cite{Sha,Cha}.

We have essentially constructed a realistic catalog of early-type systems by
simultaneously assigning stellar mass ($M_{\star}$) and stellar velocity 
dispersion ($\sigma$) to halo virial mass ($M_{\rm vir}$) in the way that the 
empirical $M_{\star}$-$\sigma$ and $M_{\rm vir}-M_{\star}$ relations are preserved. 
This catalog is not fully empirical
because $M_{\rm vir}$ is drawn from dissipationless N-body simulations rather 
than observations.  We refer to this catalog as a semi-empirical catalog (SEC).
We emphasize that the dissipationless N-body prediction of halo mass function
is robust under the $\Lambda$CDM paradigm as it is based only on gravitational
physics. Abundance matching of the halo mass function with galaxy properties 
such as stellar mass has already produced meaningful results in the literature
including the average relation between $M_\star$ and $M_{\rm vir}$ and its 
evolution with redshift and the clustering of galaxies at different redshifts 
\cite{Kra,VO,Con,Mos,Beh}. The galaxy with
$M_{\star}$ and  $\sigma$ assigned can be further assigned its stellar mass 
density profile through their observed correlations with the effective radius 
$R_{\rm e}$ and the S\'{e}rsic index $n$. These correlations are described
in appendix~A. 
For this SEC, however, dark matter density profiles are missing. 
We turn next to dynamical modeling of the systems in the SEC.

\section{Jeans dynamical modeling of early-type galaxy-halo systems}

\subsection{The spherical Jeans equation}

 For the early-type systems in the SEC described in section~3.2
 we can perform Jeans dynamical modeling to constrain the dark matter 
density profile of the embedding halo because the stellar velocity dispersion 
depends not only on the stellar mass distribution but also on the  dark matter 
distribution. We focus on the spherical Jeans equation \cite{BT} given by 
\begin{equation}
\frac{d[\rho_{\star}(r) \sigma_{\rm r}^2(r)]}{dr} 
+ 2 \frac{\beta(r)}{r} [\rho_{\star}(r) \sigma_{\rm r}^2(r)]
= - G \frac{\rho_{\star}(r) M(r)}{r^2},
\end{equation}
where $\sigma_{\rm r}(r)$ is the radial stellar velocity dispersion at radius 
$r$, $\rho_{\star}(r)$ is the three-dimensional stellar mass density at $r$, 
$M(r)$ is the total (i.e.\ stellar plus dark) mass enclosed within $r$, i.e.\ 
$M(r)=M_{\star}(r)+M_{\rm dm}(r)$, and $\beta(r)$
is the velocity dispersion anisotropy at $r$ given by
\begin{equation}
\beta(r)=1 - \frac{\sigma_{\theta}^2(r)+\sigma_{\phi}^2(r)}{2\sigma_{\rm r}^2(r)},
\end{equation}
where $\sigma_{\theta}(r)$ and $\sigma_{\phi}(r)$ are the tangential velocity 
dispersions in spherical coordinates. An integral solution of the Jeans 
equation for $\sigma_{\rm r}(r)$ is given in appendix~B.

The LOSVD of stars at projected radius $R$ on the sky $\sigma_{\rm los}(R)$ is 
given by
\begin{equation}
\sigma_{\rm los}^2(R)=\frac{1}{\Sigma_{\star}(R)} \int_{R^2}^{\infty}
\rho_{\star}(r) \sigma_{\rm r}^2(r) \left[ 1 - \frac{R^2}{r^2} \beta(r) \right]
 \frac{dr^2}{\sqrt{r^2-R^2}},
\end{equation}
where $\Sigma_{\star}(R)$ is the two-dimensional stellar mass density projected
on the sky. The stellar mass weighted $j$-th power of LOSVD within an aperture 
of radius $R$ is then given by
\begin{equation}
 \langle \sigma_{\rm los}^j \rangle (R) =
\frac{\int_0^R \Sigma_{\star}(R') \sigma_{\rm los}^j(R') R' dR'}
{\int_0^R \Sigma_{\star}(R') R' dR'}.
\end{equation}
Finally, assuming that stellar mass follows luminosity the quantity to match
 the SDSS velocity dispersion $\sigma$ is given by 
\begin{equation}
\sigma = \langle \sigma_{\rm los} \rangle (R=R_{\rm{e}}/8). 
\end{equation}

\subsection{Models for the dark matter distribution}

Initially all mass is supposed to follow a density profile similar to the
NFW profile \cite{NFW,Nav,Mer}, as predicted by N-body 
simulations.
As the galaxy is formed and settled at the center of the halo, the separate 
stellar mass distribution is embedded in the remaining dark matter distribution.
While the stellar mass distribution can be represented by the S\'{e}rsic profile
as indicated by observations, the remaining dark matter distribution is 
completely unknown. 

With the assumption that the dark matter distribution
has been smoothly readjusted from the initial NFW(-like) profile,
we model the unknown dark matter distribution using a generalized
NFW density profile, referred to as $\alpha$NFW throughout, given by 
\begin{equation}
\rho_{\alpha{\rm NFW}}(r)=\frac{\rho_{\rm s}} {\left(r/r_{\rm s}\right)^\alpha 
\left(1+r/r_{\rm s}\right)^{3-\alpha}},
\end{equation}
where $\alpha$ is the inner density power-law slope ($\alpha=1$
being the NFW value), $r_{\rm s}$ is the scale radius related to the 
concentration parameter $c_{\rm vir}$ via
\begin{equation}
c_{\rm vir} = \frac{r_{\rm vir}}{r_{\rm s}},
\end{equation}
and the parameter $\rho_{\rm s}$ is related to the dark matter mass within the
virial radius $M_{\rm dm}(=M_{\rm vir}-M_{\star})$ via
\begin{equation}
\rho_{\rm s}= \frac{M_{\rm dm}}{4 \pi r_{\rm s}^3 f_{\alpha}(c_{\rm vir})},
\end{equation}
where the function $f_{\alpha}(x)$ is given by
\begin{equation}
f_{\alpha}(x) = \int_0^x \frac{t^{2-\alpha}}{(1+t)^{3-\alpha}} dt.
\end{equation}
We also consider the Einasto profile \cite{Ein}, that is suggested by 
relatively more recent dissipationless N-body simulations \cite{Nav,Mer}, 
given by
\begin{equation}
\rho_{\rm Ein}(r)= \rho_{-2} 
\exp\left\{-(2/\tilde{\alpha})\left[(r/r_{-2})^{\tilde{\alpha}}-1\right]\right\},
\end{equation}
where $r_{-2}$ is the radius at which the logarithmic slope of the density is 
$-2$ and from which we define another concentration parameter given by 
\begin{equation}
\tilde{c}_{\rm vir} = \frac{r_{\rm vir}}{r_{-2}},
\end{equation}
and the parameter $\rho_{-2}$ is related to the dark matter mass via
\begin{equation}
\rho_{-2}=\frac{M_{\rm dm}}
         {4\pi r_{-2}^3 \tilde{f}_{\tilde{\alpha}}(\tilde{c}_{\rm vir})},
\end{equation}
where the function $\tilde{f}_{\tilde{\alpha}}(x)$ is given by
\begin{equation}
\tilde{f}_{\tilde{\alpha}}(x) = \frac{1}{\tilde{\alpha}} 
    \left(\frac{2}{\tilde{\alpha}}\right)^{-3/\tilde{\alpha}}
    \exp\left(\frac{2}{\tilde{\alpha}}\right)
    \gamma\left(\frac{3}{\tilde{\alpha}},
     \frac{2}{\tilde{\alpha}}x^{\tilde{\alpha}}\right)
\end{equation}
where $\gamma(a,b)$ is the incomplete gamma function.

\subsection{Results on dark matter density profiles}

\subsubsection{Restricted results based only on the SDSS velocity dispersion}

 For a system in the SEC with $M_{\rm vir}$ and the other parameters assigned 
the dark matter distribution given by the $\alpha$NFW (or Einasto) profile is 
completely specified by two parameters $\alpha$ and $c_{\rm vir}$
(or, $\tilde{\alpha}$ and $\tilde{c}_{\rm vir}$).  Without 
additional information the Jeans equation can be used only to solve for one of
the two parameters $\alpha$ and $c_{\rm vir}$ when the other and
  the velocity dispersion anisotropy $\beta(r)$ are specified in advance.
This can be done by equating the theoretical velocity dispersion given by 
equation~(4.5) to the SDSS velocity dispersion.
We consider the case of constant anisotropy $\beta$ and assign its value using
\begin{equation}
\beta = \left\{ \begin{array}{cl} x & \mbox{\hspace{1em} if $0 \le x < 0.5$} \\
      1.8x & \mbox{\hspace{1em} if $-0.5 < x < 0$}  \end{array}, \right. 
\end{equation}
where $x$ is an uniform random variable and we are using the prior
$-0.9 < \beta < 0.5$ from stellar dynamics analyses of handfuls of early-type 
galaxies in the literature \cite{Ger,Cap2,GS}. We further consider the case
that either $\alpha=1$ or $c_{\rm vir}$ is fixed at the Bolshoi predicted
value for all halos (i.e.\ the weighted mean for distinct halos and subhalos).
In some cases, for a given value of $\beta$ and the rest of the  
parameters ($M_\star$, $R_{\rm e}$, $\cdots$) assigned already to the system there
may not exist a solution for equation~4.5. In this case another value of
$\beta$ is drawn and retried until a solution is found. In some rare cases
this effort fails even for some significant number of trials. In such a case
we remove the system from the catalog (i.e.\ reject the whole parameter set
assigned). This means that the posterior distribution of $\beta$ can be
different from the prior constraint.

\begin{figure}
\begin{center}
\setlength{\unitlength}{1cm}
\begin{picture}(15,11)(0,0)
\put(-0.7,12.){\includegraphics{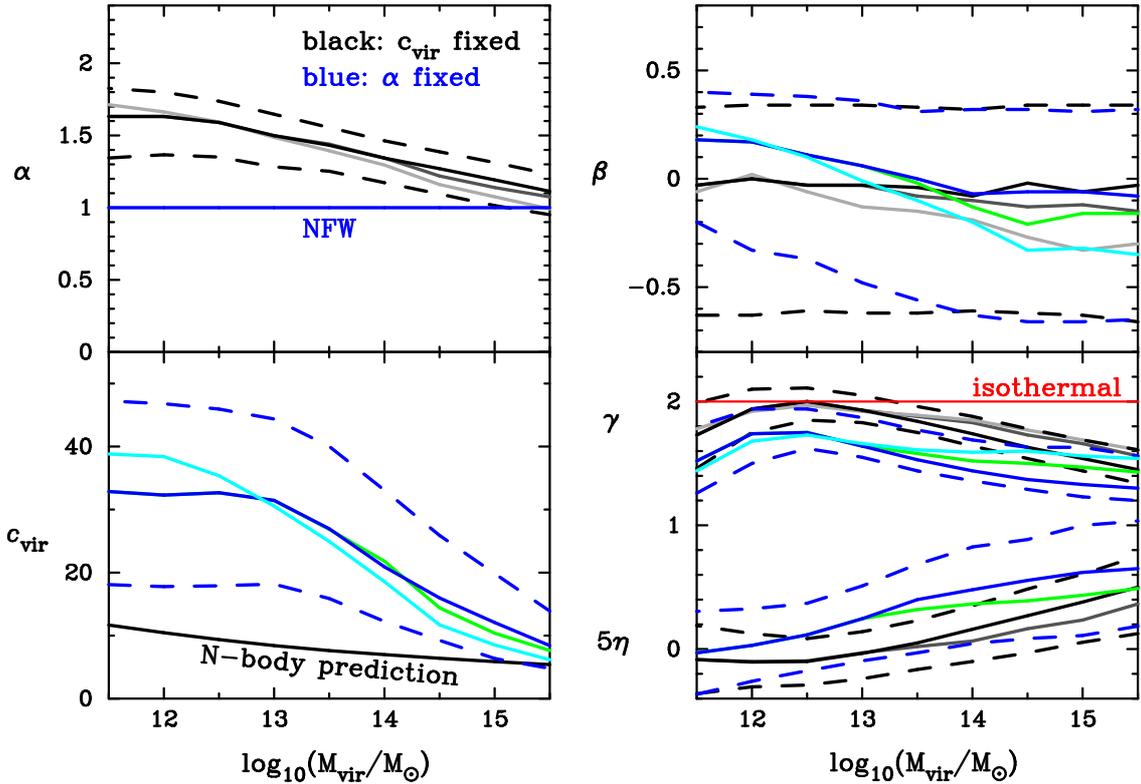}}
\end{picture}
\caption{
(Left) Distributions of the $\alpha$NFW halo dark matter density 
profile parameters $\alpha$ and $c_{\rm vir}$ with either of them fixed 
obtained through the spherical Jeans equation based on 
SDSS velocity dispersions. 
Solid curves are the median 
 while dashed curves are the 68\% probability limits. 
The dark gray and green curves are based on corrections of $M_\star$ and 
$\sigma$ due to possible systematic errors of abundance matching for
clusters while
the light gray and cyan curves include further corrections of stellar masses 
due to a possible systematic variation of IMF as a function of $\sigma$.
(Right) Upper panel shows the simultaneously 
obtained distribution of
constant anisotropy $\beta$ with the prior $[-0.9,0.5]$. 
Lower panel shows the minus logarithmic slope $\gamma$ of
the predicted total (i.e.\ stellar plus dark matter) mass density profile and
the logarithmic slope $\eta$ (multiplied by 5 for visibility) of 
the velocity dispersion profile at $r = R_{\rm e}/2$.
}
\label{fix2}
\end{center}
\end{figure}

Fig.~\ref{fix2} shows the results for the $\alpha$NFW profile as a function of 
 $M_{\rm vir}$. For the first case $c_{\rm vir}$ is fixed 
at the value predicted by the Bolshoi N-body simulation. 
In this case the resulting values of $\alpha$ lie above the NFW 
value $\alpha=1$ with increasingly larger values as $M_{\rm vir}$ 
decreases. For the second case,  $\alpha$ is fixed at the NFW value and 
$c_{\rm vir}$ is determined by the Jeans equation. Similarly to the first case 
the values of $c_{\rm vir}$ lie above the N-body simulation prediction.
These results indicate that dark matter densities are boosted in the inner 
halos of early-type galaxies. 
 The density boost is greater for a lower mass halo. 
Qualitatively similar results are obtained for the Einasto profile as well.
If $\tilde{\alpha}$ is fixed at $0.17$ (the N-body predicted value) \cite{Nav},
 the resulting values of $\tilde{c}_{\rm vir}$ tend
 to lie above the Bolshoi prediction. If $\tilde{c}_{\rm vir}$ is fixed
at the value predicted by the Bolshoi simulation, the resulting values of 
$\tilde{\alpha}$ tend to lie below $\tilde{\alpha}=0.17$ implying steepened
inner density slopes.

Stellar masses used above are based only on the Chabrier initial 
mass function (IMF) of stars \cite{Ber}.
Recently, there are reports of observational studies that show systematic 
variations of IMFs across galaxy populations depending on 
velocity dispersion \cite{Treu}, stellar mass-to-light ratio \cite{Cap12},
or color \cite{Dut12}. To quantify systematic effects due to IMFs
we consider a systematically varying IMF with velocity
dispersion by which stellar masses based on the Chabrier IMF are multiplied
by $1.7 \times 10^{1.31 \log_{\rm 10}(\sigma/{\rm{km}}~{\rm{s}}^{-1}) - 3.14}$ \cite{Treu}
with a constant scatter of $0.15$~dex. Furthermore, as shown in Fig.~\ref{biv} 
our statistical match between galaxies and halos gives
 a lower $M_\star$ by up to $0.2$~dex at fixed $M_{\rm vir}$ for 
$M_{\rm vir}> 10^{13} M_{\odot}$ compared with other independent results 
that are based on IMFs similar to the Chabrier. 
 To gauge possible systematic effects due to stellar mass 
uncertainties we obtain alternative results taking into account these stellar 
mass corrections/adjustments. The correction of $0.2$~dex in $M_\star$ due
to abundance matching is accompanied by a corresponding correction in $\sigma$
so as to preserve the observed $M_\star$-$\sigma$ relation.
The alternative results are shown in Fig.~\ref{fix2}.
 For the case of using the fixed NFW inner slope $\alpha=1$
 ${c}_{\rm vir}$ is lowered by up to $\sim 40$ percent for massive clusters
  when both systematic effects are applied (cyan curves).

The above two cases (fixing $\alpha$/$\tilde{\alpha}$ or 
$c_{\rm vir}$/$\tilde{c}_{\rm vir}$)
 imply different inner density profiles that cannot be
distinguished by the  SDSS velocity dispersion alone. 
It shows the degeneracy of the allowed density profiles given only the
 parameters of a system in the SEC described above (section~3.2). 
In fact, the degeneracy is broader than the above two cases. 
In other words, other combinations of 
$\alpha$ and $c_{\rm vir}$ (or, $\tilde{\alpha}$ and $\tilde{c}_{\rm vir}$) 
can satisfy the Jeans equation as well. To break the 
degeneracy we need additional constraints. Those constraints may also help to
constrain the velocity dispersion anisotropy. 

The upper right panel of Fig.~\ref{fix2} shows the posterior 
distributions of constant velocity dispersion anisotropy $\beta$.
In some cases (mostly the cases of fixing $\alpha=1$) the posterior 
distributions are clearly different from the prior
distribution given by equation~(4.14). As mentioned above,
this occurs because only certain combinations of 
$\{\alpha/\tilde{\alpha}, c_{\rm vir}/\tilde{c}_{\rm vir}, \beta\}$
can satisfy the Jeans equation for a system whose other parameters
have been specified in advance and a randomly selected $\beta$ may or may not 
be a solution.

The bottom right panel of Fig.~\ref{fix2}  shows the 
distributions of the negative logarithmic slope $\gamma$ of the 
three-dimensional total radial density profile at $R_{\rm e}/2$, i.e.\  
$\gamma \equiv -d\ln[\rho_{\star}(r)+\rho_{\rm dm}(r)]/d\ln r$.
The two cases give the values of $\gamma$ offset by $\approx 0.2-0.3$.
Fig.~\ref{fix2} also shows the distributions of
the logarithmic slope $\eta$ (see section~4.3.2)
of the velocity dispersion profile at $R_{\rm e}/2$.
There is also an offset of $\approx 0.02-0.07$ in $\eta$ between the two cases.
The distributions of $\gamma$ and $\eta$ in the two cases shown in 
Fig.~\ref{fix2} hint that dark matter density boost in the inner halo
 is more consistent with enhanced $\alpha$ at fixed $c_{\rm vir}$ than
enhanced $c_{\rm vir}$ at fixed $\alpha$ because the former case gives 
distributions of $\gamma$ and $\eta$ that are more consistent with current
gravitational lensing \cite{Rus,Koo,Bar} and spectroscopic \cite{Cap,Jor}
 observations of early-type galaxies. We consider below (section~4.3.2)
incorporating spectroscopic observations into our SEC through Jeans modeling.

\subsubsection{Dark matter density profiles statistically matched with velocity dispersion profiles: completion of the semi-empirical catalog}

The SDSS velocity dispersion corresponds to the luminosity-weighted LOSVD 
within $R_{\rm e}/8$. Clearly, this alone cannot constrain uniquely the
dark matter distribution in the halo. Measured velocity dispersions at multiple 
radii for each SDSS galaxy would be most useful in breaking the degeneracy. 
Without them we consider independent observational constraints on the velocity
dispersion profile (VP) of early-type galaxies \cite{Cap,Jor}.  
The available observational constraints show that the logarithmic slope
 $\eta$ of the luminosity-weighted LOSVD profile at $R_{\rm e}/2$, i.e.\  
$\eta \equiv d\ln\langle\sigma_{\rm los}\rangle (R)/d\ln R$, ranges from 
$-0.066 \pm 0.035$ \cite{Cap} to $-0.04$ \cite{Jor}.
We assume that the mean value of $\eta$ is given by the mean of these two
independent measurements, 
i.e., $\langle \eta \rangle = -0.053$. For the dispersion of $\eta$ we 
assume the root-mean-square of the measured dispersion $0.035$ for one sample
 and the difference 
between the two independent means $0.026$, i.e., $\sigma_\eta = 0.044$.

We then proceed as follows. For each system in the SEC 
 we first obtain a degenerate set of $\{\alpha,c_{\rm vir},\beta \}$. 
We do this by assigning random values to $\alpha$  and $\beta$ and then 
solving for $c_{\rm vir}$ using equation~(4.5). We assume the priors
$0.1 <\alpha< 2.5$ and equation~(4.14) for $\alpha$ and $\beta$ respectively.
We then select one out of this degenerate set for each system using a selection
 function $S(\eta)$ so that the posterior distribution of $\eta$ for
all systems in the SEC matches the observed distribution described above. 
An uniform selection function would result in a posterior distribution of 
$\eta$ that is biased toward a shallower VP (i.e.\ $\eta > -0.53$).
After numerical experiments we choose the following selection function 
\begin{equation}
S(x) \propto \left\{ \begin{array}{cl} 
\left[1-\theta(x-\mu)\right] 
         \exp\left[ -\frac{(x-\mu)^2}{2 \sigma_{\rm L}^2} \right] +
      \theta(x-\mu) \exp\left[ -\frac{(x-\mu)^2}{2 \sigma_{\rm H}^2} \right]
  & \mbox{\hspace{1em} if $ x > -0.2$} \\
      0 & \mbox{\hspace{1em} else}  \end{array}, \right. 
\end{equation}
where $\mu =-0.066$, $\sigma_{\rm L}=0.1$, 
$\sigma_{\rm H}=0.01$ and $\theta(x)$ is the theta function. 
Notice that the above selection function has the peculiar property that
the most probable value is displaced downward and the upward width is 
negligibly small. Any selection function having these features would do the
same job. After selecting an $x$ using this selection function 
we select one set $\{\alpha,c_{\rm vir},\beta \}$ (out of the above degenerate 
set for the given system in the SEC) that best matches $x$
 by minimizing the figure-of-merit function defined by
\begin{equation}
Q^2_{\rm (VP)} \equiv \left(\frac{\eta-x}
{\sigma_\eta} \right)^2.
\end{equation} 
If the value of $Q^2_{\rm (VP)}$ is greater than $13.7$ for the best-fit set, 
then the whole degenerate set is rejected. This criterion matters only for
$M_{\rm vir} \gtrsim 10^{13.5} {\rm M}_{\odot}$  and is chosen so that the posterior
distribution of $\eta$ deviates minimally from our adopted observational 
constraint of $\eta = -0.053 \pm 0.044$  while not causing any bias in the
distributions of other parameters in the SEC. A criterion lower 
(i.e.\ stronger) than $13.7$ would bring the posterior distribution of $\eta$ 
closer to the observational constraint but the resulting systems tend to have 
biased distributions of $R_{\rm e}$ and $n$. A weaker criterion would allow
$\eta$ to deviate more. Our assumption is the minimal deviation. Because of
this difficulty our procedure becomes less reliable as $M_{\rm vir}$ increases
toward massive clusters.

\begin{figure}
\begin{center}
\setlength{\unitlength}{1cm}
\begin{picture}(13,9)(0,0)
\put(-1.,9.){\includegraphics{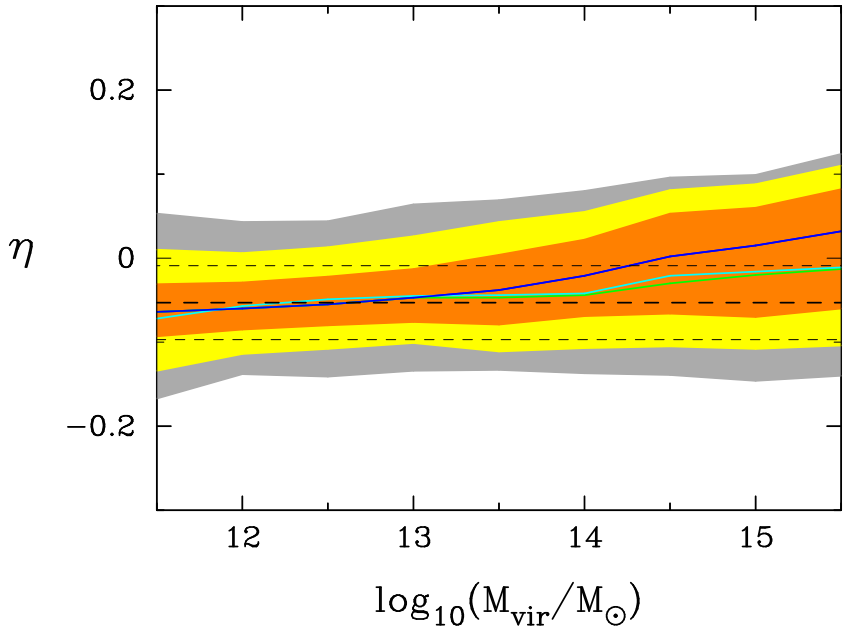}}
\end{picture}
\caption{
The posterior distribution of $\eta$ (the VP slope at 
$R_{\rm e}/2$) for our early-type systems is compared with the input 
observational constraint of $\eta = -0.053 \pm 0.044$ (black dashed lines).
 The blue solid curve is the expectation value while
the orange, yellow and gray regions contain respectively 68\%, 95\%, and 99.7\%
of the systems. The green and cyan curves are for the corrected stellar masses
as in Fig.~\ref{fix2}. These results are based on Jeans dynamical modeling of
our systems in the SEC using the observational constraint on $\eta$ as a
statistical input.
}
\label{VP}
\end{center}
\end{figure}

Fig.~\ref{VP} shows the posterior distribution of $\eta$. The mean value
agrees well with the input empirical mean for 
$M_{\rm vir} \lesssim 10^{13.5} {\rm M}_{\odot}$ but deviates upward systematically
as $M_{\rm vir}$ increases. As mentioned above this deviation is unavoidable
without biasing the distributions of observed galaxy parameters such as 
$R_{\rm e}$. In this sens our results require the VP slope to vary systematically
with $M_{\rm vir}$. This deviation does not necessarily mean a discrepancy 
between our results and the observed VPs because the latter \cite{Cap,Jor} 
cannot be used to address any systematic variation that may be present. 
The galaxy-to-galaxy dispersion of $\eta$ is somewhat lower than the adopted 
$\sigma_\eta = 0.044$ for $M_{\rm vir}\lesssim 10^{13.5}{\rm M}_{\odot}$
but higher for $M_{\rm vir}\gtrsim 10^{13.5}{\rm M}_{\odot}$. Note, however, that
$\sigma_\eta = 0.044$ includes a systematic error between the two samples used
(see above). Our dispersion is more consistent with the intrinsic scatter
of $0.035$ in one sample \cite{Cap}. 

Through the above procedure we have effectively assigned  dark matter density
profile, velocity dispersion profile slope at $R_{\rm e}/2$ and  constant
velocity dispersion anisotropy simultaneously to each system in the SEC
 so that the Jeans equation is satisfied along with all the adopted
observational constraints. We have thus determined all the unknowns of the
SEC making it a complete catalog of early-type galaxy-halo systems.

\begin{figure}
\begin{center}
\setlength{\unitlength}{1cm}
\begin{picture}(15,11)(0,0)
\put(-0.7,12.){\includegraphics{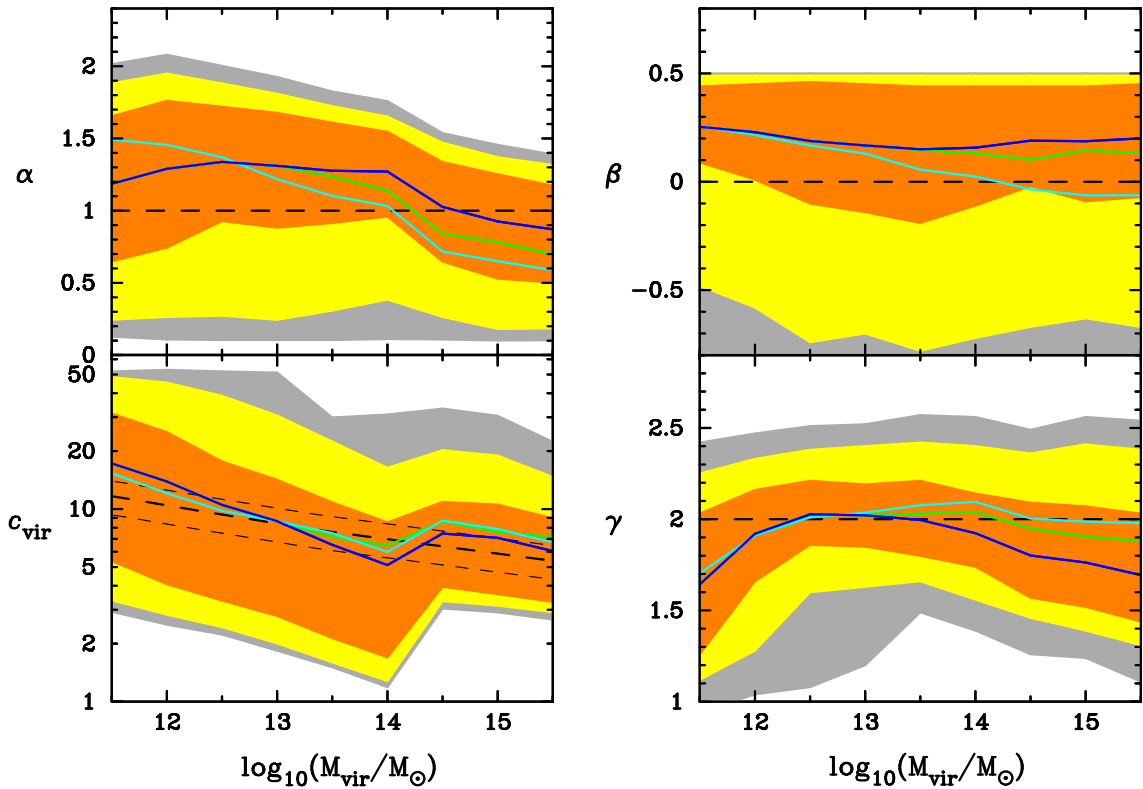}}
\end{picture}
\caption{
Distributions of the same parameters of Fig.~\ref{fix2}, i.e.\ 
$\alpha$ and $c_{\rm vir}$, in our final SEC for which each system is assigned 
a velocity dispersion profile satisfying the spherical Jeans equation. 
The blue solid curves show the expectation values. 
The green and cyan solid
curves are for the corrected stellar masses as in Fig.~\ref{fix2}.
Orange, yellow and gray regions contain the 68\%, 95\% and 99.7\% 
occurrences respectively. 
The black dashed line in the lower left panel shows the mean NFW
concentration predicted by the Bolshoi N-body simulation with
thin dashed lines showing a systematic error of 20\%.  
The black dashed 
line in the lower right panel corresponds to the isothermal profile.
}
\label{prof}
\end{center}
\end{figure}

 Fig.~\ref{prof} shows the distribution of the dark matter density
 profiles based on the $\alpha$NFW model in the final SEC.
The distributions of $\alpha$ and $c_{\rm vir}$ show 
that their expectation values are varying systematically with $M_{\rm vir}$ 
and there are significant halo-to-halo variations at fixed $M_{\rm vir}$.
The expectation values of $\alpha$ lie above the NFW value for
$M_{\rm vir} \lesssim 10^{13.5-14.5}$~M$_{\odot}$ and are typically 
$\langle \alpha \rangle \approx 1.3$ for  galactic halos with 
$10^{12} {\rm M}_{\odot} \lesssim M_{\rm vir} \lesssim 10^{13-14} {\rm M}_{\odot}$. 
The halo-to-halo root mean square (rms) scatter is typically 
${\rm rms}(\alpha) \sim 0.4-0.5$. Hence for our $\sim 1000$ halos
at fixed $M_{\rm vir}$ the estimated error of $\langle \alpha \rangle$ is
$\sim 0.015$ implying that $\langle\alpha\rangle \approx 1.3$ for 
$10^{12}{\rm M}_{\odot}\lesssim M_{\rm vir}\lesssim 10^{13-14} {\rm M}_{\odot}$ is 
some $20\sigma$ above the NFW value $\alpha=1$. The NFW profile is clearly
ruled out as a mean profile of galactic-scale halos implying significant 
effects of halo contraction. 
The trend of $c_{\rm vir}$ follows overall the N-body prediction \cite{Kly,Pra}.
The expectation values are within 20\% (systematic error) 
of the N-body prediction except for
low mass galactic halos with $M_{\rm vir} \lesssim 10^{12} {\rm M}_{\odot}$ for
which the expectation values are clearly higher.

Fig.~\ref{prof} also shows the distribution of the negative logarithmic slope 
$\gamma$ at $R_{\rm e}/2$ for the total mass distribution. For halos with 
$10^{12}~{\rm M}_{\odot} \lesssim M_{\rm vir} \lesssim 10^{14}~{\rm M}_{\odot}$ 
the values of  $\gamma$ scatter around
the isothermal value $\gamma=2$ with the expectation value of 
$1.9 \lesssim \langle\gamma\rangle \lesssim 2.1$ and the rms galaxy-to-galaxy
scatter of ${\rm rms}(\alpha)\approx 0.2-0.3$.
 This result for a general sample of early-type galaxies is 
in good agreement with the results for strong lensing observations of
early-type galaxies \cite{Rus,Koo,Bar}.
 For massive cluster-sized halos the central total density profile within 
the optical region tends to be shallower than the isothermal 
 for stellar masses based on the Chabrier IMF but becomes close 
to the isothermal for stellar masses based on a systematic variation of IMF 
as a function of $\sigma$.
The distribution of  velocity dispersion anisotropy $\beta$ is tilted toward
a mild radial anisotropy with the expectation value close to isotropy.
This distribution is consistent with independent results for
dozens of individual systems
\cite{Ger,Cap2,GS}.

\begin{figure}
\begin{center}
\setlength{\unitlength}{1cm}
\begin{picture}(15,11)(0,0)
\put(-0.7,12.){\includegraphics{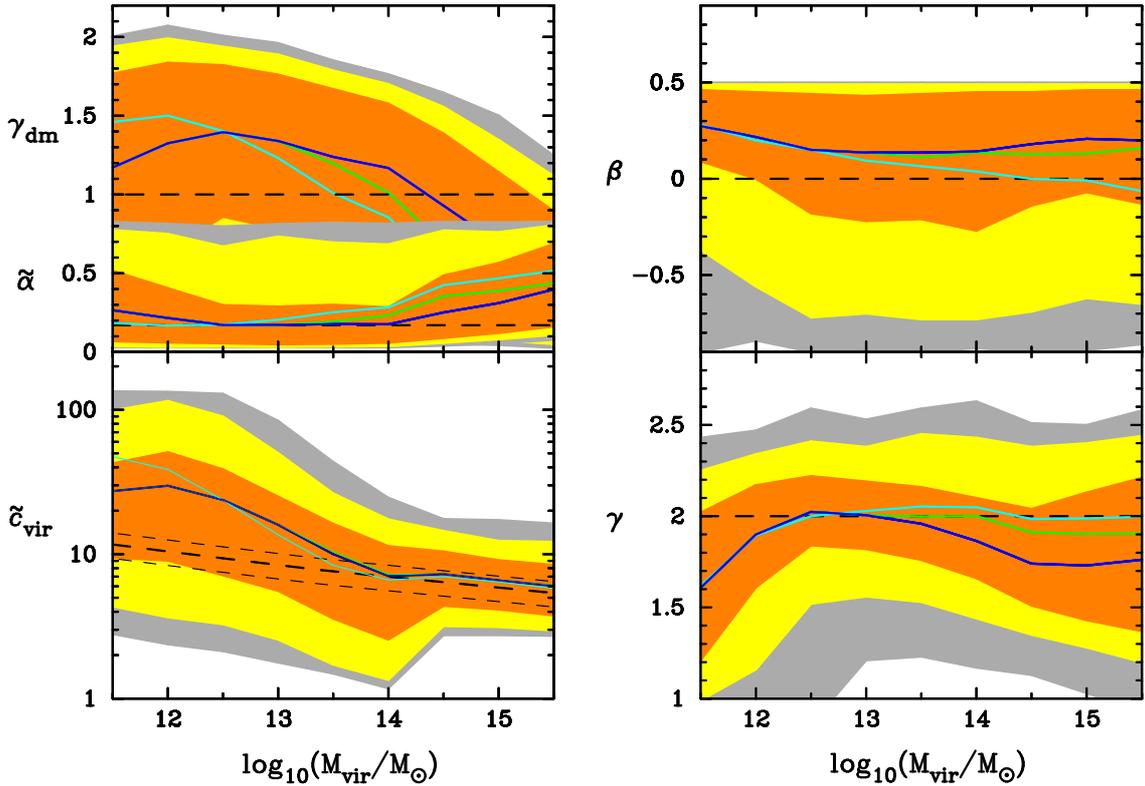}}
\end{picture}
\caption{
Same as Fig.~\ref{prof} except that the Einasto model parameters 
$\tilde{\alpha}$ and $\tilde{c}_{\rm vir}$ are shown. Parameter
$\gamma_{\rm dm}$ is the minus logarithmic slope for the Einasto model
dark matter distribution, i.e.\ 
$\gamma_{\rm dm} \equiv - d \ln \rho_{\rm Ein}(r)/d \ln r$ at $R_{\rm e}/2$. 
The constrained values of $\gamma_{\rm dm}$ match well with those of
$\alpha$ of the $\alpha$NFW model shown in Fig.~\ref{prof}.
}
\label{profEin}
\end{center}
\end{figure}

Fig.~\ref{profEin} shows the corresponding results for the Einasto model. 
The distributions of $\beta$ and $\gamma$ are strikingly similar 
to those for the $\alpha$NFW model. The distributions of 
 $\tilde{\alpha}$ and $\tilde{c}_{\rm vir}$ cannot be directly compared with 
those of $\alpha$ and $c_{\rm vir}$. For the Einasto model the logarithmic slope 
of the density varies continuously toward the origin, i.e.\
 $\gamma_{\rm dm}=2(\tilde{c}_{\rm vir} r/r_{\rm vir})^{\tilde{\alpha}}$
for $\rho_{\rm dm}(r) \propto r^{-\gamma_{\rm dm}}$, so that unlike the $\alpha$NFW
model an inner density slope cannot be characterized by a single parameter. 
At a fixed inner radius the slope $\gamma_{\rm dm}$ can be boosted either 
by increasing $\tilde{c}_{\rm vir}$  or decreasing $\tilde{\alpha}$. 
Fig.~\ref{profEin} indicates that the former is more likely the 
case for galactic halos. 
 Namely, for $M_{\rm vir}\lesssim 10^{13-14}{\rm M}_{\odot}$ 
$\langle\tilde{\alpha}\rangle \sim 0.17$ with 
$\langle\tilde{c}_{\rm vir}\rangle$ greater than the N-body prediction.
However, for cluster halos $\langle\tilde{c}_{\rm vir}\rangle$ is within 20\%
of the N-body prediction while $\langle\alpha\rangle$ tends to
deviate from $0.17$.
The distribution of $\gamma_{\rm dm}$ at 
$R_{\rm e}/2$ is similar to that of $\alpha$ of the $\alpha$NFW model. 

We have tested our  systems against a couple of dynamical 
mass scaling relations 
presented recently in the literature. One scaling involves
the mass within $r \sim R_{\rm e}/2$ and the LOSVD at that radius \cite{Chu}
while the other scaling involves the mass within $r \approx 4 R_{\rm e} /3$  
and the luminosity-weighted LOSVD to infinity \cite{Wol}. 
For our systems the average masses predicted by these scalings 
are accurate within 10\% for  
$10^{12}~{\rm M}_{\odot} \lesssim M_{\rm vir} \lesssim 10^{14}~{\rm M}_{\odot}$ 
and there are system-to-system variations with a typical dispersion of 
5\% to 10\% at fixed $M_{\rm vir}$ (appendix~D). 

\section{Implication for baryon-induced halo contraction and characterization of the contracted profiles}

The distribution of dark matter density profiles 
 in our final semi-empirical catalog (SEC) 
of early-type systems supports the hypothesis 
that halos are modified in response to dissipational gas cooling
and galaxy formation \cite{Blu,Gne04}. 
Our results indicate that  galactic halos with 
$M_{\rm vir} \lesssim 10^{13.5}$~M$_{\odot}$, on average, have contracted  
and dark matter density profiles of the contracted halos are diverse. 
Fig.~\ref{fix2} and Fig.~\ref{prof} show that halo contraction is more likely 
to be realized by steepening of inner density slope ($\alpha$) rather than 
decrease of scale radius (i.e.\ increase of concentration $c_{\rm vir}$) as
 the former is more consistent with the current observational constraints on
the velocity dispersion profile of early-type galaxies.

 To quantify halo contraction we consider
 the ratios of the dark matter masses of our halos to 
those of NFW halos with the average concentrations predicted by the Bolshoi 
dissipationless N-body simulation within three radii, 
i.e.\ the effective radius $R_{\rm e}$, the NFW scale 
radius $r_{\rm s,NFW}$ and one fifth of the halo virial radius ($r_{\rm vir}/5$).
If the NFW profile were assumed for the initial N-body predicted halo and  
there were no modification to the halo due to baryonic effects, then
these ratios should center around unity with scatters arising solely from 
the scatters of N-body predicted NFW concentrations.

\begin{figure}
\begin{center}
\setlength{\unitlength}{1cm}
\begin{picture}(15,11)(0,0)
\put(-0.7,12.){\includegraphics{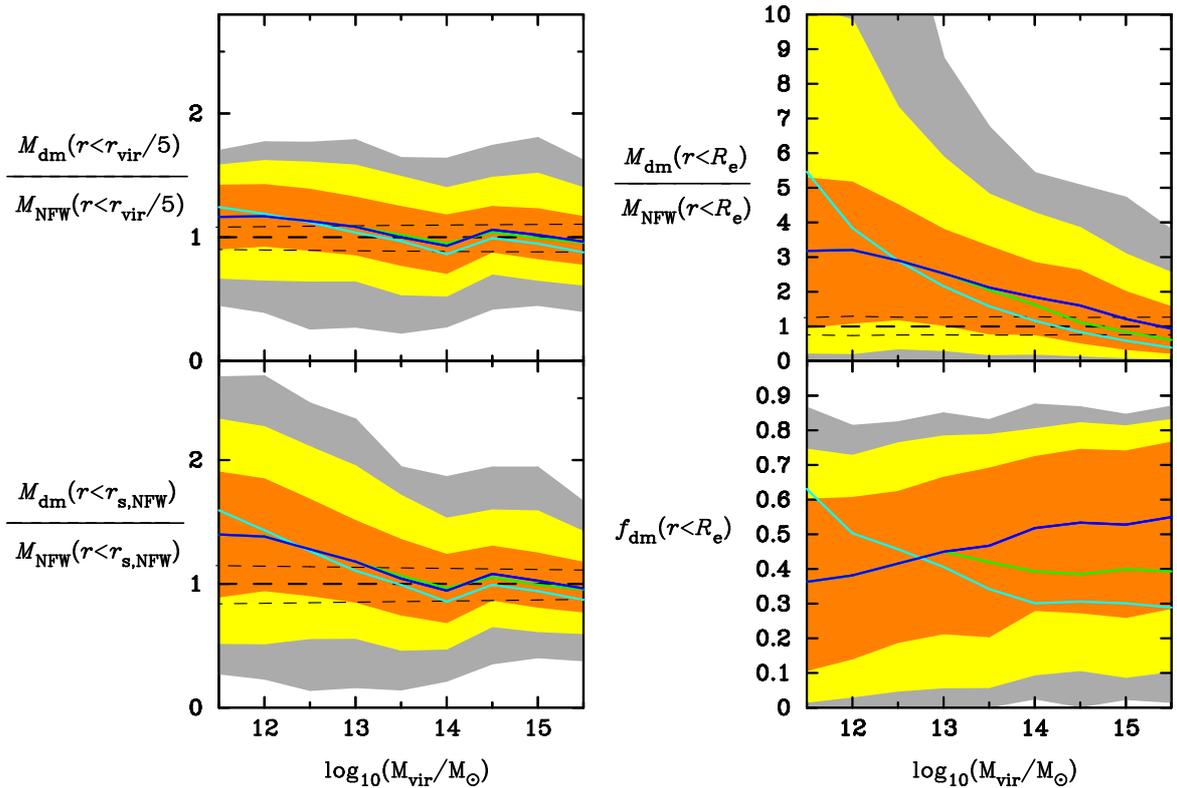}}
\end{picture}
\caption{Dark matter masses and fractions based on our Jeans 
dynamical modeling results for the $\alpha$NFW model. Color representations 
of lines and regions are the same as in Fig.~\ref{prof}. The corresponding 
results for the Einasto model are similar to those shown here.
(Left panels \& Top Right)
Ratio of the dark matter masses within a certain three dimensional radius for
the initial galaxy-less NFW halo ($M_{\rm NFW}$) and the 
realistic halo embedding an early-type galaxy at the center ($M_{\rm dm}$). 
Here $r_{\rm s,NFW}$ is the scale radius of the NFW profile while $R_{\rm e}$
is the effective radius of the embedded central galaxy. 
The black thick dashed line is the expected mean (i.e.\ unity) for the case
of no modification of the halo while thin dashed lines show 20\% systematic
error in the N-body predicted $c_{\rm vir}$.
 These results show that most halos have contracted in the optical region of 
central galaxies with the varying degree 
of mass enhancement depending on the the halo virial mass. However, the 
contraction has modified the halo up to the scale radius only for relatively 
less massive halos with $M_{\rm vir} \lesssim 10^{13} {\rm M}_{\odot}$ and 
starts to be
unimportant beyond the scale of 20\% of the virial radius. (Bottom Right)
Fraction of dark matter mass within $R_{\rm e}$ of the early-type 
galaxy embedded at the center of a halo.
}
\label{Mrat}
\end{center}
\end{figure}

 The calculated ratios are displayed in Fig.~\ref{Mrat}.
It shows evidently dark matter density enhancement 
in the optical region of the inner halo. The mean density 
enhancement factor within $r=R_{\rm e}$ increases steeply from $\approx 1$ 
(no enhancement)  at $M_{\rm vir} \approx 10^{15-15.5}$~M$_{\odot}$ 
(or $10^{13.5-14}$~M$_{\odot}$) up to 3 -- 4 at 
$M_{\rm vir}\approx 10^{12}$~M$_{\odot}$. As $M_{\rm vir}$ decreases, however, 
the scatter also increases rapidly indicating greater diversity.
 The Kolmogorov-Smirnov (K-S) tests of the distributions or 
the Student t tests of the means at fixed $M_{\rm vir}$ rule out the null
hypothesis that halo profiles follow the dissipationless N-body
predictions with a typical dispersion of $0.14$~dex in $c_{\rm vir}$ \cite{Mac} 
even after allowing for a systematic error of $20\%$ in the mean concentrations
(K-S and t probabilities $\lesssim 10^{-5}$).
The K-S and t tests show that our halo profiles can be consistent with
the N-body predictions at $M_{\rm vir} \approx 10^{15-15.5}$~M$_{\odot}$ if the
Chabrier IMF is assumed but at $10^{13.5-14}$~M$_{\odot}$ if the IMF varies
systematically with stellar velocity dispersion \cite{Treu}.
Interestingly, the dark matter mass within $r=R_{\rm e}$ for our halos
 based on the systematically varying IMF is lower than the N-body prediction
for $M_{\rm vir} \gtrsim 10^{14.5}$~M$_{\odot}$ suggesting a possibility of inner 
halo expansion in those massive clusters. However, this possibility of halo
expansion for massive clusters should be taken with caution as our procedure
is less reliable for massive clusters (see section~4.3.2) due to uncertainties
in velocity dispersion profiles and stellar masses.

At the scale of $r_{\rm s,NFW}$ (i.e.\ the N-body predicted scale radius assuming
the NFW model)  the halo density enhancement is weakly present only 
 for $M_{\rm vir} \lesssim 10^{13}$~M$_{\odot}$. At the scale of $r_{\rm vir}/5$ 
 the mean density enhancement is $\approx 20\%$ at best for galactic halos.
A statistical analysis of weak lensing effects for $r \ge r_{\rm vir}/5$ based on
statistically representative samples of galaxies, groups and clusters from SDSS 
yields concentrations similar to N-body predictions assuming the NFW profile 
for $10^{12} {\rm M}_{\odot} \lesssim M_{\rm vir} \lesssim 10^{15} {\rm M}_{\odot}$ 
\cite{Man2}. This result is consistent with our finding of no or little density
 enhancement within $r = r_{\rm vir}/5$. The weak lensing measured scatter of 
concentration gives rise to $1\sigma$ scatter of 10\% -- 20\% in mass within 
$r = r_{\rm vir}/5$ consistent with our result shown in Fig.~\ref{Mrat}.

Fig.~\ref{Mrat} also shows the fraction of dark matter mass within a sphere 
 of radius $r=R_{\rm e}$ for our halos embedding early-type galaxies. 
The mean dark matter fraction increases with $M_{\rm vir}$ for the
case of Chabier IMF but decreases for the case of the systematically varying
IMF. The mean fraction lies 0.2 to 0.5 for 
$10^{12}~{\rm M}_{\odot} \lesssim M_{\rm vir} \lesssim 10^{14}~{\rm M}_{\odot}$
depending on  $M_{\rm vir}$ and the assumption of IMF.
Dark matter fraction at fixed $M_{\rm vir}$
covers a broad range showing a great diversity.
Dynamical analyses of nearby elliptical galaxies indicate $\sim 30$\%
of dark matter contribution within $R_{\rm e}$ \cite{Cap2} and a recent analysis
of strong lensing galaxies indicates that $f_{\rm dm}(r<R_{\rm e})$ can be
$\sim 0.2$ -- 0.6 \cite{Bar}. These results agree well with our results.

Traditionally, adiabatic contraction has been invoked for the  response of
 halos to dissipational baryonic processes \cite{Blu,Gne04,Duf,Gne11}. 
Hydrodynamic simulations have been used to study adiabatic contraction in detail
recently \cite{Duf,Gne11}. Independent of hydrodynamic simulations 
and relying only on SDSS observed galaxy data and robust classical results from 
N-body dissipationless simulations, our results demonstrate evidently that halo 
contraction takes place in most occurrences of spheroidal galaxy formation.  

Contraction of halo mass profile in response to the baryon condensation by 
itself is expected to lead to steepening of the central density profile out to
 $\sim 5 - 10\%$ of the virial radius but leave the outer mass profile intact 
\cite{Gne04}. 
However, hierarchical formation of central galaxy may also result in 
an additional mild increase of halo concentration due to several other 
processes \cite{Rud}.
Our results are in line with these expectations.

An observational indication of adiabatic contraction in a statistically
representative sample of SDSS elliptical 
galaxies is reported recently \cite{Sch} based on a combination of weak 
lensing measurements of halos and stellar velocity dispersions. These authors
assume the NFW profile and notice that their analysis requires higher 
concentrations of halos than pure dark matter simulation halos. 
A combined analysis of strong and weak lensing effects for 28 clusters 
from the Sloan Giant Arcs Survey finds increasingly over-concentration 
as halo mass gets lower assuming the NFW profile \cite{Ogu}. 

Based on generalized dark matter density profiles 
that are allowed to vary from the NFW profile, our results 
provide not only (semi-empirical) evidence for halo contraction but also a 
characterization of contraction. Our results indicate that contracted halo 
profiles generally deviate from the NFW and display a range of diversity and 
halo contraction is more frequently realized by steepening of the radial 
density slope rather than rescaling of the NFW profile. If the NFW 
profile were forcibly used to describe the contracted halo, then the 
concentration would be significantly higher than that of the pure dark matter 
simulation halo which would match high concentrations of halos obtained by 
recent observational studies assuming the NFW profile \cite{Sch,Ogu}. 

Our analysis presented here relies on N-body simulation data as well as
observed galaxy data. In this sense our results are referred to as 
semi-empirical.  It would be of great importance to carry out a similar but
fully empirical analysis that relies only on empirical information. 
A development of such an analysis is under way and will be presented in the
near future. However, our current analysis has the advantage that we can 
naturally construct a mock universe into a N-body simulation lightcone. 
Such a mock universe will be particularly useful for realistic strong lensing 
simulation for future surveys 
because baryonic effect is a crucial factor for strong lensing. 

It would be useful to carry out a semi-empirical (as was done here) or fully 
empirical analysis for other morphological types of galaxies. 
An analysis based on empirical information (satellite kinematics and weak 
lensing observations of halos and observed galaxy scaling relations) and 
N-body simulation results indicates that halos may expand in response to 
baryonic effects for late-type systems \cite{Dut}.
 Recently, $\Lambda$CDM N-body + SPH (smoothed particle hydrodynamic) 
simulations of low-mass systems with $M_{\rm vir}\lesssim 10^{11.5}{\rm M}_{\odot}$
($M_{\star} \lesssim 10^{9.4}{\rm M}_{\odot}$) show that supernovae driven outflows
can transfer enough energy to dark matter particles so that dark halo expands
transforming initially cuspy central profile into a shallower core \cite{Gov}.
Another recent cosmological hydrodynamic simulation of a massive spiral galaxy 
with $M_{\rm vir}\approx 10^{12}{\rm M}_{\odot}$ also argues for halo expansion 
\cite{Mac12}. 

 The literature results of halo expansion for late-type and dwarf systems
contrast strikingly with our results of halo contraction for early-type systems.
This apparent contrast is already noticed and discussed in the literature 
\cite{Dut}. If this contrast is real (i.e.\ both expansion and contraction 
results are correct), merging driven galaxy transformation \cite{TT}
 may play the vital role in creating contracted halos of early-type galaxies. 
Also, our finding that the degree of halo contraction for early-type galaxies 
is a decreasing function of halo mass indicates that the effects of merging
depend on various factors such as morphological types of merging subunits, 
the amount of gas (i.e.\ whether wet or dry) and perhaps AGN 
 driven outflows from the central supermassive black holes. 
 However, before interpreting halo expansion/contraction it is more important
to address thoroughly possible systematic errors of our and literature results.
Possible systematic errors of our results include the distribution of stellar
 IMFs, velocity dispersion anisotropy shapes, velocity dispersion profiles,
the N-body simulation produced halo mass function, and the spherical mass 
models (the real early-type systems are ellipsoidal in general).
 We have addressed the first three of these possible errors here although
they can be (and should be) better addressed with better empirical information 
in the future. The use of abundance matching has also been checked against
empirical halo mass-stellar mass relations and its possible systematic error
for large halo mass has been addressed. However, we have not yet addressed 
the effect of non-sphericity of mass models. An optimistic prediction is
that oblate-like and prolate-like shapes are equally likely so that their
effects cancel statistically. To quantify the effect of non-sphericity 
non-spherical Jeans equations need to be used in the future based on 
empirical information on shapes. With regard to hydrodynamic simulation results
for halo expansion in the literature empirical methods should also be used in 
the future to test those results independently and (more importantly) 
empirically.
 
\section{Characterizing dark matter annihilation strength in the halos of early-type galaxies}

One proposed way of identifying dark matter particles has been the detection of 
$\gamma$-ray emission from dark matter annihilation or decay in the halos
\cite{Por,Ull}.
In the case of dark matter annihilation the predicted $\gamma$-ray flux is 
known to be very sensitive to dark matter density. Specifically,
the predicted $\gamma$-ray flux from dark matter annihilation is proportional to
the dark matter density squared integrated along the line-of-sight.
 When averaged over the angular scale of the halo it can be approximated by
\cite{Ack}
\begin{equation}
J_{\rm dm} \simeq \frac{1}{D^2} \int_{\rm Vol} r^2 \rho_{\rm dm}^2(r) dr,
\end{equation}
where $D$ is the distance to the halo. 

\begin{figure}
\begin{center}
\setlength{\unitlength}{1cm}
\begin{picture}(15,15)(0,0)
\put(0.,-0.5){\includegraphics{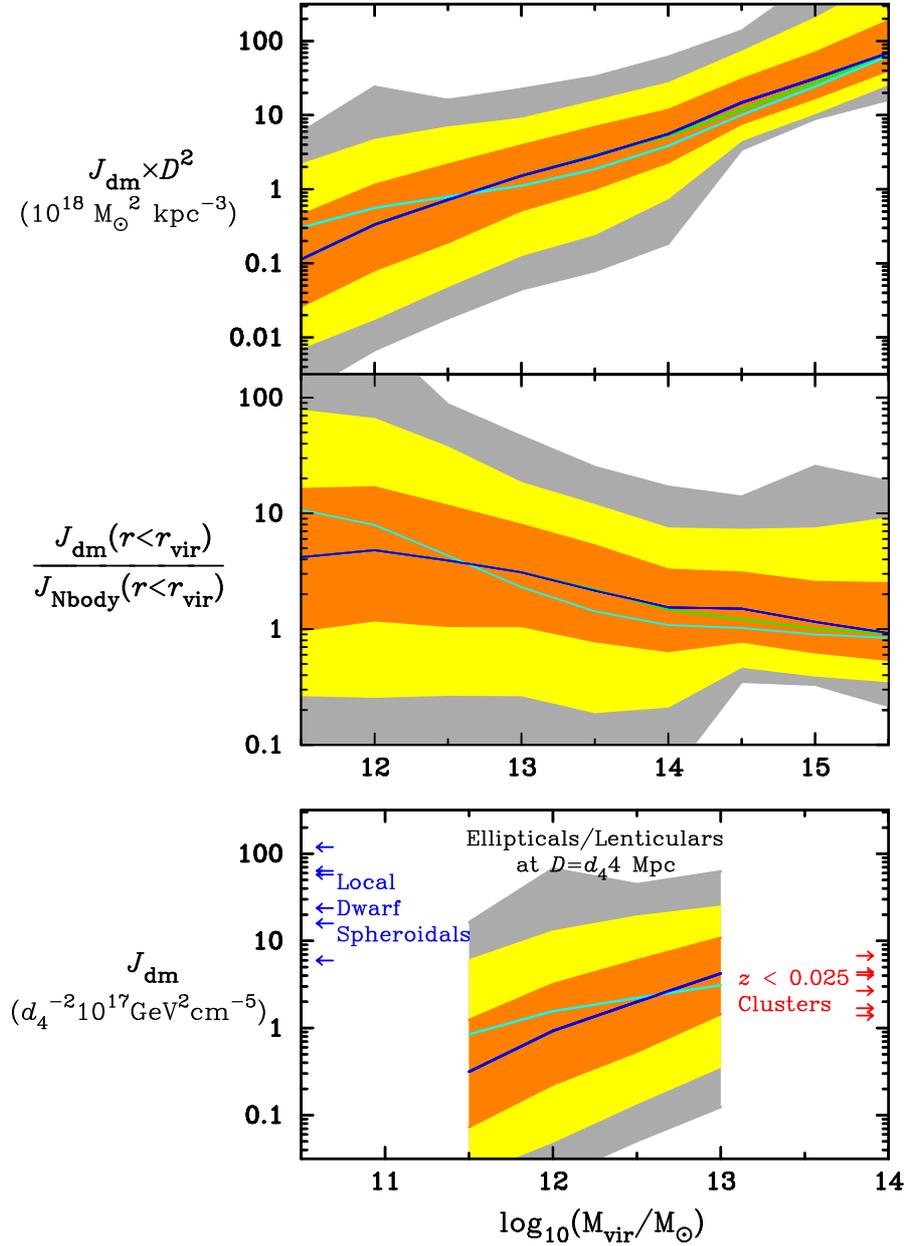}}
\end{picture}
\caption{
(Top) The dark matter density squared integrated along the 
line-of-sight and averaged on the halo angular scale, $J_{\rm dm}$  times the 
halo distance squared $D^2$, approximated by 
$J_{\rm dm} D^2 \simeq \int r^2 \rho^2(r)dr$ \cite{Ack} 
for the halos in our final SEC. 
The results are shown only for the Einasto model because
 $J_{\rm dm}$ diverges for certain central cusps of the $\alpha$NFW model.
The predicted $\gamma$-ray flux from dark matter annihilation
is proportional to $J_{\rm dm}$. 
(Middle) Ratio of $J_{\rm dm}$ for the halo embedding a central
early-type galaxy to that for the dark matter only N-body
simulation halo with $\tilde{\alpha}=0.17$ and
 $\tilde{c}_{\rm vir}$ matched to the Bolshoi concentrations. This shows that the
 predicted $\gamma$-ray flux is enhanced in most galactic halos embedding 
spheroidal galaxies. 
(Bottom) The predicted range of $J_{\rm dm}$ for nearby normal spheroidals 
(i.e.\ ellipticals and lenticulars) at a fiducial distance of $D=d_4 4$~Mpc is
shown and compared with the estimates for local dwarf spheroidal 
galaxies (blue arrows) \cite{Abd} and low-redshift 
($z<0.025$) clusters (red arrows) \cite{Ack}.
}
\label{J}
\end{center}
\end{figure}

Clearly, halo contraction is expected to boost the predicted $\gamma$-ray flux
 from dark matter annihilation.  Fig.~\ref{J} shows the predicted values of
$J_{\rm dm}\times D^2 $ for our constrained halo dark matter distributions based
 on the Einasto model. The $\alpha$NFW model cannot be used for calculating
$J_{\rm dm}$ because $J_{\rm dm}$  diverges for $\alpha > 1.5$ as $r\rightarrow 0$.
The specific values of $J_{\rm dm}$ at a fiducial distance of $D=d_4 4$~Mpc are
also shown and compared against several local dwarf
spheroidal galaxies \cite{Abd} and low redshift clusters \cite{Ack} that 
have been  proposed as promising targets for $\gamma$-ray flux. 
Fig.~\ref{J} suggests that typical nearby early-type galaxies  with 
$M_{\rm vir} \sim 10^{12}$~M$_{\odot}$ -- $10^{13}$~M$_{\odot}$ can also be
 promising targets. 
This rather surprising prediction is the consequence of the mass dependent
 enhancement of $J_{\rm dm}$ as shown in the middle panel of Fig.~\ref{J},  
resulting from the varying degree of halo contraction.  
We note that there are two well-known elliptical or lenticular galaxies 
at a distance of $\sim 4$~Mpc, Centaurus~A (NGC~5128) and Maffei~1.
Although our results are only for the halos embedding early-type galaxies at
their centers, they highlight the importance of halo contraction induced by
galaxy formation for dark matter search.

\section{Conclusions}

We have constructed a semi-empirical catalog of early-type galaxy-halo systems
through a combination of the observed statistical properties of SDSS galaxies 
and the halo mass function from the Bolshoi N-body dissipationless simulation
in conjunction with the observed properties of velocity dispersion profiles of 
early-type galaxies. In constructing the catalog we have 
 determined dark matter density profiles through Jeans dynamical
modeling. Based on the systems in our catalog we find the following statistical
properties for early-type galaxies and their embedding halos:

\begin{itemize}

\item The distribution of total mass density profiles within the 
effective radius at fixed $M_{\rm vir}$ scatters around the isothermal profile
for $10^{12} {\rm M}_{\odot} \lesssim M_{\rm vir} \lesssim 10^{14} {\rm M}_{\odot}$.
The density slope $\gamma$ at $r=R_{\rm e}/2$ for $\rho(r)\propto r^{-\gamma}$
has a mean value $1.9 \lesssim \langle \gamma \rangle \lesssim 2.1$ with a 
rms scatter of $\approx 0.2-0.3$.

\item The inferred dark matter density profiles of the halos imply significantly
 higher dark matter densities in the inner regions compared with those 
from the dissipationless simulation. The mean density boost factor
within a sphere of $R_{\rm e}$ ranges from
$\approx 1$ for  $M_{\rm vir} \approx 10^{15-15.5} {\rm M}_{\odot}$
(or $10^{13.5-14} {\rm M}_{\odot}$)  to $\approx 3-4$ for 
$M_{\rm vir} \approx 10^{12} {\rm M}_{\odot}$.
This provides an independent support for
halo contraction for galactic halos.

\item The inferred mean dark matter densities within a sphere of 
$r_{\rm vir}/5$ are at most $\approx 20$\% higher than the N-body prediction 
implying that the outer mass profiles are minimally affected by halo 
contraction.

\item Using a couple of general-class models for the dark matter distribution 
we obtain statistical characterizations of the contracted profiles. 
For galactic halos with 
 $10^{12} {\rm M}_{\odot} \lesssim M_{\rm vir} \lesssim 10^{13} {\rm M}_{\odot}$,
the three-dimensional dark matter density slope $\gamma_{\rm dm}$ at 
$r=R_{\rm e}/2$ for $\rho_{\rm dm}(r)\propto r^{-\gamma_{\rm dm}}$ has a mean value
of $1.2 \lesssim \langle \gamma_{\rm dm} \rangle \lesssim 1.4$ with a 
rms scatter of $\approx 0.4-0.5$. The NFW profile is clearly ruled
out as a mean profile although some fraction of halos may well follow it.

\item The dark matter fraction within the sphere of radius 
$R_{\rm e}$ has a mean value ranging from 0.2 to 0.5 with a typical rms 
scatter of  $\approx 0.2-0.3$ for 
$10^{12} {\rm M}_{\odot} \lesssim M_{\rm vir} \lesssim 10^{15} {\rm M}_{\odot}$.

\item Halo contraction boosts significantly dark matter annihilation 
strength in the halos embedding early-type galaxies so that nearby early-type
galaxies may be promising targets for indirect dark matter search.

\end{itemize}

\section*{Acknowledgments}

 We would like to thank Dan Hooper for conversations and comments on the 
draft and Anatoly Klypin for making outputs from the Bolshoi and Multidark 
simulations available to us. 
We would also like to thank the anonymous referees
for useful suggestions, Aaron Dutton and Michele Cappellari for 
comments/discussions on stellar IMFs.
 KHC is grateful for the sabbatical leave 
(March 2010 -- January 2011) at Fermilab Center for Particle Astrophysics 
where significant parts of this work were carried out.
He thanks in particular Albert Stebbins and Craig Hogan at FCPA for their 
encouragements. He also thanks In-Taek Gong at Sejong University for his 
assistance in preparing a figure. AVK was supported in part by the NSF grant 
AST-0708154, and by the Kavli Institute for Cosmological Physics at the 
University of Chicago through the NSF grant PHY-0551142 and PHY-1125897 and an
endowment from the Kavli Foundation. MB is grateful for partial support 
provided by NASA grant ADP/NNX09AD02G.

\appendix

\section{SDSS early-type galaxy parameter correlations}

An early-type galaxy with $M_{\star}$ and $\sigma$ specified may be given its 
stellar mass
density profile using observed correlations with $R_{\rm e}$ (effective radius)
 and $n$ (S\'{e}rsic index) assuming the  S\'{e}rsic
 stellar mass density profile \cite{Ser} given by
\begin{equation}
\Sigma_{\star}(X) = A_n \exp\left( -b_n X^{1/n} \right), 
\end{equation}
where $X=R/R_{\rm e}$ ($R$ being the two dimensional radius), 
$b_n = 2n-1/3+0.009876/n$ ($0.5 < n < 10$), and 
$A_n = b_n^{2n}/[2\pi n \Gamma(2n)]$ \cite{PS,Mar}. 
This is the projected two dimensional density normalized such that its 
integrated total mass is unity. 
The deprojected three dimensional density at $x=r/R_{\rm e}$ (the three 
dimensional radius in units of $R_{\rm e}$) is then given by
\begin{equation}
\rho_{\star}(x) = \rho_n x^{-\alpha_n} \exp\left( -b_n x^{1/n} \right) 
\end{equation}
with $\alpha_n=1-0.6097/n+0.05563/n^2$ and 
$\rho_n=(b_n)^{n(3-\alpha_n)}/\{4\pi n \Gamma[n(3-\alpha_n)]\}$ \cite{PS,Mar}.

\begin{figure}
\begin{center}
\setlength{\unitlength}{1cm}
\begin{picture}(12,15)(0,0)
\put(-0.5,-1.2){\includegraphics{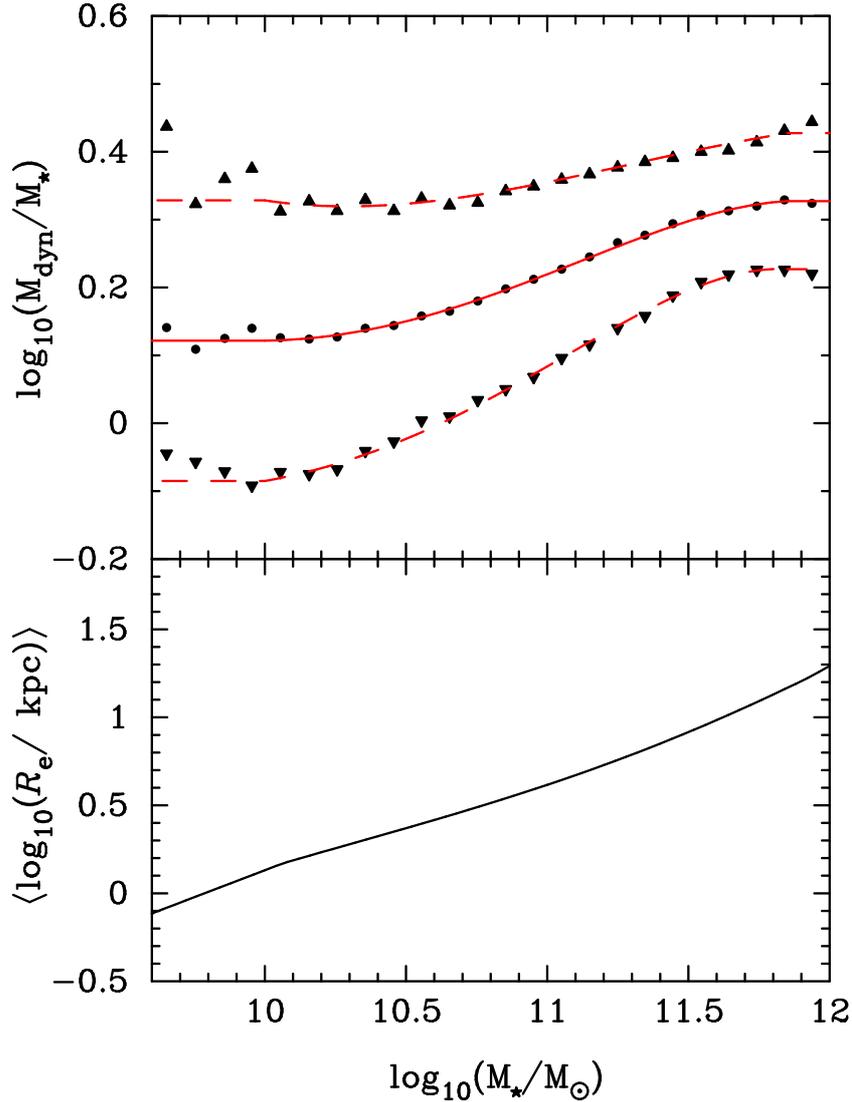}}
\end{picture} 
\caption{
(Top) The distribution of $\log_{10}(M_{\rm dyn}/M_\star)$ for SDSS early-type
 galaxies with $M_{\rm dyn} \equiv 5R_{\rm e}\sigma^2/G$.
We obtain this distribution from the data analyzed in \cite{Ber}.
Data points are the median values and 68\% scatters. Red solid curve and 
dashed curves are the polynomial fits assuming the Gaussian distribution. 
(Bottom) The distribution of the mean value of
 $\log_{10}(R_{\rm e}/{\rm kpc})$ based on the Gaussian mean of 
$\log_{10}(M_{\rm dyn}/M_\star)$ from the top panel and the mean of 
$\log_{10}(\sigma/{\rm km}~{\rm s}^{-1})$ from Fig.~\ref{MsVD}. 
}
\label{MdynMs}
\end{center}
\end{figure}

To a galaxy with $M_{\star}$ and $\sigma$ specified we give a specific value of
$R_{\rm e}$ using the observed distribution of $M_{\rm dyn}/M_{\star}$ as a 
function of $M_{\star}$, where $M_{\rm dyn} \equiv 5R_{\rm e}\sigma^2/G$ 
($G$ being Newton's gravitational constant) is a `dynamical' mass.
Fig.~\ref{MdynMs} shows the observed distribution of $M_{\rm dyn}/M_{\star}$. 
Notice that the distribution of $M_{\rm dyn}/M_{\star}$ essentially shows the
scatter of the fundamental mass plane relation of early-type galaxies
for three parameters $R_{\rm e}$, $\sigma$ and $M_{\star}$.

\begin{figure}
\begin{center}
\setlength{\unitlength}{1cm}
\begin{picture}(10,12)(0,0)
\put(-0.5,-1.){\includegraphics{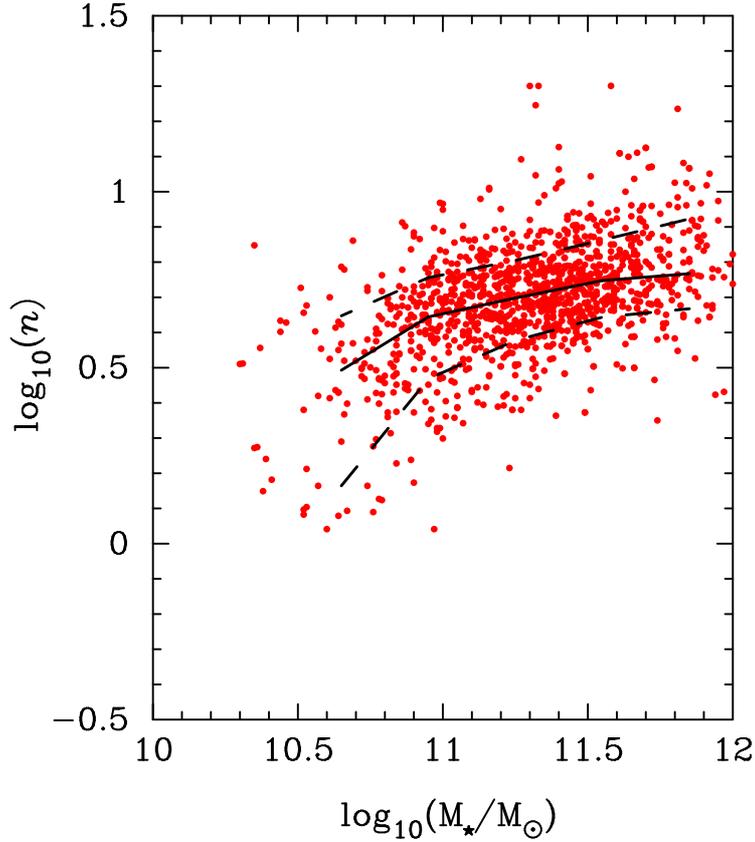}}
\end{picture} 
\caption{
The distribution of S\'{e}rsic index ($n$) as a function of stellar mass 
($M_\star$) for SDSS early-type galaxies (red points) with the curves showing 
median values (full) and 68\% scatters (dashed). 
These data are taken from \cite{Guo}.
}
\label{nser}
\end{center}
\end{figure}

Finally, we give a specific value of $n$ to a galaxy using the observed
distribution of $n$ as a function of $M_{\star}$ (see Fig.~\ref{nser}). 
This result comes also from SDSS data \cite{Guo}. S\'{e}rsic index $n$ 
may also be correlated with $\sigma$ and $R_{\rm e}$. 
However, the current unavailability of such correlations 
forces us to use only the correlation shown in Fig.~\ref{nser}.

\section{An integral solution of the Spherical Jeans equation}

For a specific form of the velocity dispersion anisotropy $\beta(r)$ the radial
velocity dispersion $\sigma_{\rm r}(r)$ can be expressed as an integral form 
that can be easily evaluated numerically. 

\begin{figure}
\begin{center}
\setlength{\unitlength}{1cm}
\begin{picture}(15,11)(0,0)
\put(-0.8,11.5){\includegraphics{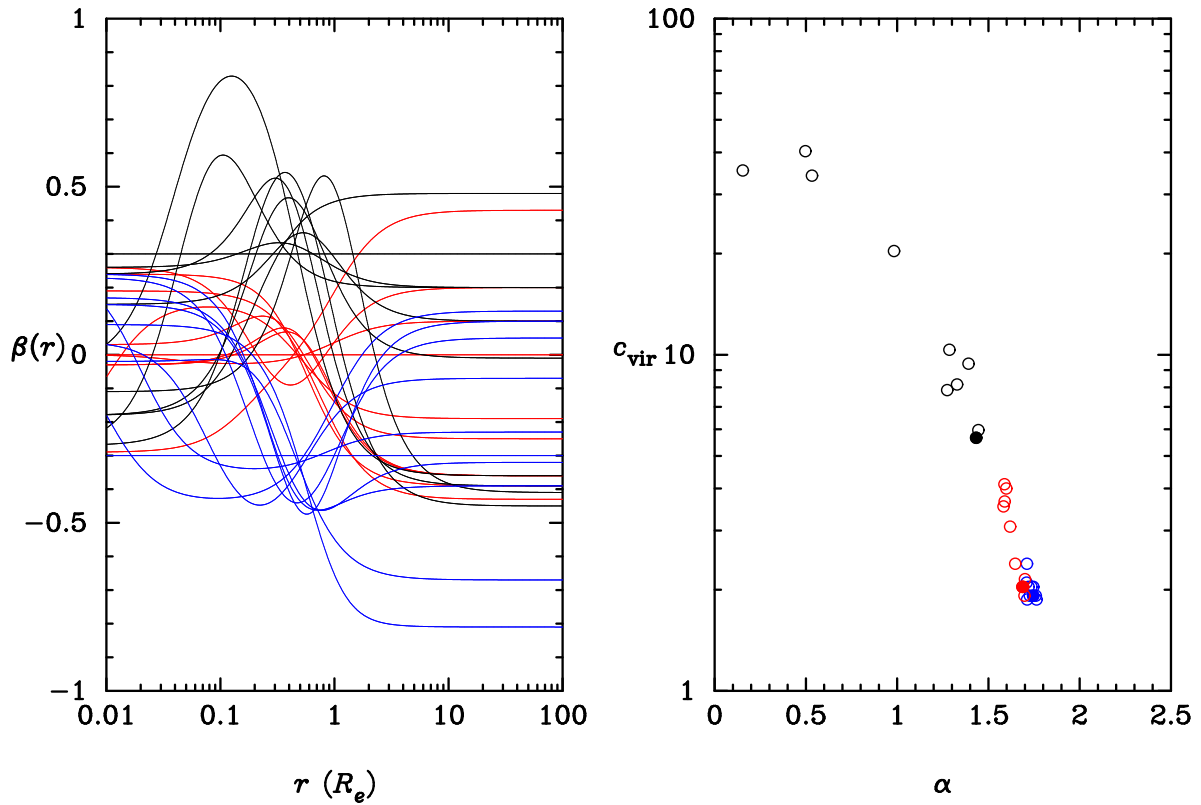}}
\end{picture} 
\caption{
(Left) Examples of varying anisotropy $\beta(r)$ based on the model given by 
equation~(B3) that solve for the spherical Jeans equation for a system with
$M_{\rm vir}=10^{13} {\rm M}_\odot$ (see the text). Black, red and blues curves are
respectively for $\beta_{\rm mean}=+0.3,0,-0.3$. (Right) Dark matter density 
profile parameters of the $\alpha$NFW model (eq.~4.6) matching the anisotropies
shown left for the same system. For $\beta_{\rm mean}=+0.3$ varying 
anisotropies (open black circles) can give significantly different values of 
$\alpha$ and $c_{\rm vir}$ compared with the case of constant anisotropy 
(filled black circle) whereas for $\beta_{\rm mean}=-0.3$ (blue) varying 
anisotropies have little effects on $\alpha$ or $c_{\rm vir}$. 
For $\beta_{\rm mean}=0$ (red), varying anisotropies can change appreciably
$c_{\rm vir}$ only. 
}
\label{btest}
\end{center}
\end{figure}

We recast the spherical Jeans equation [eq.\ (4.1)] as
\begin{equation}
 \frac{dy(r)}{dr} + p(r) y(r) = q(r),
\end{equation}
where $y(r) \equiv \rho_\star(r) \sigma_{\rm r}^2(r)$, 
$p(r) \equiv 2\beta(r)/r$, and 
$q(r) \equiv -G \rho_\star(r) M(r)/r^2$.
Then, with the definition of 
$\omega(r) \equiv \exp\left[ \int^r p(t) dt \right]$ 
we get
\begin{equation}
 y(r)=-\frac{1}{\omega(r)} \int_r^{\infty} \omega(t) q(t) dt.
\end{equation}
Let us now consider a general functional form for $\beta(r)$ given by 
\begin{equation}
\beta(r)=\beta_0 + \beta_1\frac{r^2}{r^2+r_1^2} +\beta_2\frac{r^2}{r^2+r_2^2},
\end{equation}
where $r_1 < r_2$ and $\beta_0$ is the anisotropy at the origin. This model
allows an extremum anisotropy at a finite, non-zero $r$.
We can relate $\beta_1$ and $\beta_2$ in equation~(B.3) to a mean anisotropy 
within $r_c$ ($\beta_{\rm mean}$) and the anisotropy at infinity ($\beta_\infty$)
as follows:
\begin{eqnarray}
\beta_1 &=& \frac{\beta_{\rm mean}-\beta_0 (1-u_2)-\beta_\infty u_2}{u_1 - u_2}, \\
\beta_2 &=& \beta_\infty - \beta_0 -\beta_1,
\end{eqnarray}
where $u_1 = 1 - (r_1/r_c) \arctan(r_c/r_1)$ and 
$u_2 = 1 - (r_2/r_c) \arctan(r_c/r_2)$. Fig.~\ref{btest} shows
examples of varying anisotropies based on equation~(B3) for which we
consider three cases of $\beta_{\rm mean} (r< R_{\rm e})=0, +0.3,-0.3$ and take 
 values from the ranges $-0.3 < \beta_0 < 0.3$, $-0.9 < \beta_\infty < 0.5$, 
$0 < r_1 < R_{\rm e}$ and $0 < r_2 < R_{\rm e}$ (with $r_2 > r_1$).  

For this model $\omega(r)$ becomes
\begin{equation}
\omega(r) = r^{2\beta_0} (r^2 + r_1^2)^{\beta_1}(r^2 + r_1^2)^{\beta_2}. 
\end{equation}
Substituting this into equation~(B.2) we get
\begin{equation}
\sigma^2_{\rm r}(r)=G
   \int_r^\infty \frac{\omega(t)}{\omega(r)}  
   \frac{\rho_\star(t)}{\rho_\star(r)} \frac{M(t)}{t^2} dt. 
\end{equation}
For the case of constant anisotropy 
$\beta=\beta_0=\beta_{\rm mean}=\beta_\infty$ it takes the
following simple form
\begin{equation}
\sigma^2_{\rm r}(r)=G\frac{r^{-2\beta}}
 {\rho_\star(r)}  \int_r^\infty t^{2(\beta-1)}\rho_\star(t)M(t)dt. 
\end{equation}

In the above the total mass within $r$ is given by 
$M(r) = M_\star(r) + M_{\rm dm}(r)$ with the stellar mass
\begin{equation}
M_\star(r)=M_\star \frac{\gamma[n(3-\alpha_n),b_n(r/R_{\rm e})^{1/n}]}
      {\Gamma[n(3-\alpha_n)]}
\end{equation}
and the dark matter mass
\begin{equation}
M_{\rm dm}(r)=
(M_{\rm vir}-M_\star)\frac{f_\alpha(r/r_s)}{f_\alpha(c_{\rm vir})} 
\end{equation}
for the $\alpha$NFW model and
\begin{equation}
M_{\rm dm}(r)= (M_{\rm vir}-M_\star)\frac{\tilde{f}_{\tilde{\alpha}}(r/r_{-2})}
     {\tilde{f}_{\tilde{\alpha}} (\tilde{c}_{\rm vir})}, 
\end{equation}
for the Einasto model. 
In the above $\Gamma(x)$ and $\gamma(x,y)$ are the gamma function
 and the incomplete gamma function respectively. The functions $f_\alpha(x)$ and
 $\tilde{f}_{\tilde{\alpha}}(x)$ are given by 
equation~(4.9) and equation~(4.13) respectively.

\section{Effects of varying anisotropies of velocity dispersions}

In the main text we assume that anisotropies of velocity dispersions are 
constant in radius. Constant anisotropies are used mainly because of the 
computational simplicity and in part because of the lack of empirical
statistical characterization of radial behaviors of anisotropies. Here we 
consider artificial varying anisotropies $\beta(r)$ using equation~(B.3) 
introduced above to test whether the constancy of anisotropies 
is likely to have biased our results on dark matter density profiles. 

\begin{figure}
\begin{center}
\setlength{\unitlength}{1cm}
\begin{picture}(15,11)(0,0)
\put(-0.7,12.){\includegraphics{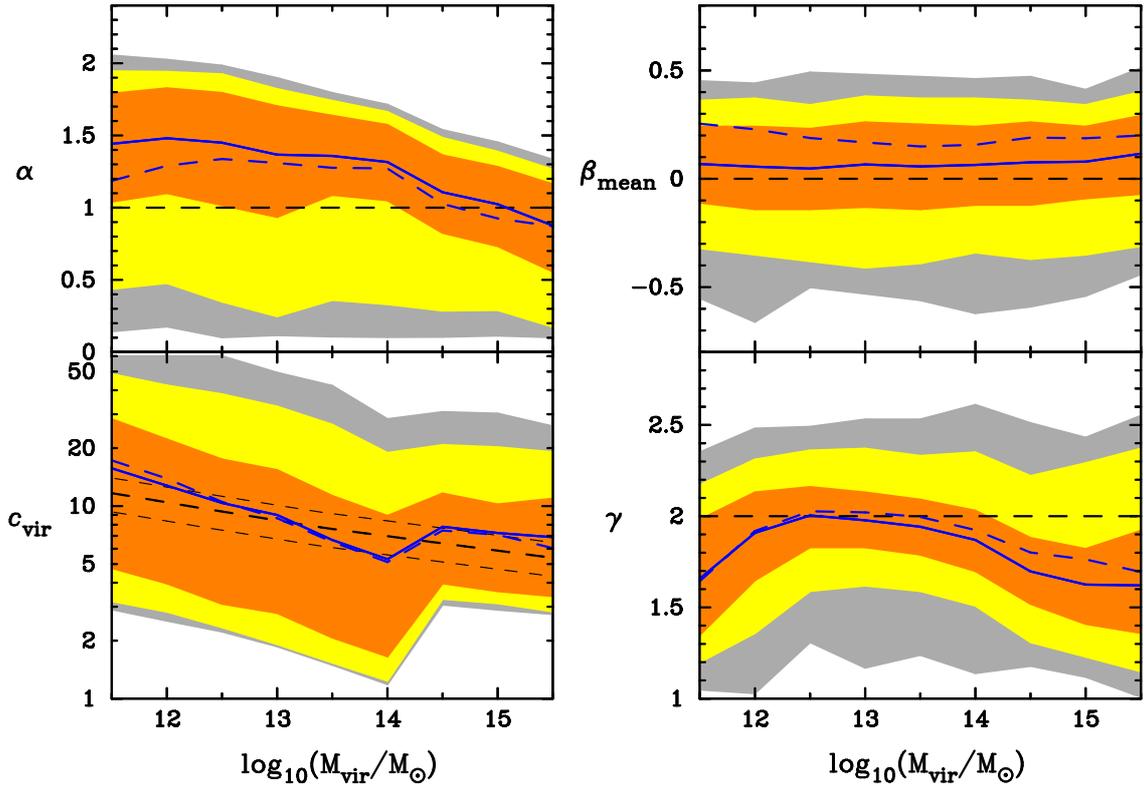}}
\end{picture} 
\caption{
Same as Fig.~\ref{prof} except that this result is based on varying anisotropies
given by equation~(B.3). Blue full and dashed curves show respectively the 
expectation values for the cases of varying and constant anisotropies. Notice
that there are only small offsets between the two.
}
\label{profbetr}
\end{center}
\end{figure}

Both observed anisotropies of stellar kinematics \cite{Ger} and simulated 
anisotropies of dark matter kinematics \cite{Nav} show that real anisotropies 
can vary significantly with $r$ in the inner region of halos. 
However, those anisotropies 
are bounded within $-0.9 \lesssim \beta(r) \lesssim 0.5$ for all probed $r$. 
In particular, $\beta(r) \sim 0$ as $r \rightarrow 0$. Using equation~(B.3)
for $\beta(r)$ we modify the procedure of the main text (section~4.3) as
follows. For each galaxy we fix $\beta_{\rm mean}$ within $R_{\rm e}$ using a 
value $x$ drawn randomly from a probability density function
  \begin{equation}
P(x) \propto 
\left[1-\theta(x-\mu)\right] 
        \exp\left[ -\frac{(x-\mu)^2}{2 \sigma_{\rm L}^2} \right] +
      \theta(x-\mu) \exp\left[ -\frac{(x-\mu)^2}{2 \sigma_{\rm H}^2} \right]
 \end{equation}
with $\mu=0.18$, $\sigma_{\rm H}=0.11$ and $\sigma_{\rm L}=0.25$ based on 
$\sim 40$ unoverlapping early-type galaxies in the literature \cite{Ger,Cap2}. 
This rather strong prior is necessary because of the great freedom allowed by
equation~(B.3) and is justified because it comes from observed galaxies.
We then vary the rest of the anisotropy parameters ($\beta_0$, $\beta_\infty$, 
$r_1$, and $r_2$) to create a degenerate model set for each system using the 
following priors: $-0.3 < \beta_0 < 0.3$, equation~(4.14) for $\beta_\infty$,
$0 < r_1 < R_{\rm e}$, and $0 < r_2 < R_{\rm e}$ with $r_1 < r_2$ (this range for
$r_1$ and $r_2$ is chosen to allow for rapidly varying in the inner region). 
These ranges are intended to encompass observed variations in early-type 
galaxies.  Fig.~\ref{profbetr} shows
the results on the dark matter density profiles based on the Chabrier stellar
masses matching Fig.~\ref{prof} with constant anisotropies.
We notice that the statistical distributions of the parameters are quite 
similar to those for the case of constant anisotropy, with only small offsets 
of the means between the two cases.

To understand the above results we make a detailed case study of a
system with $M_{\rm vir} = 10^{13}~{\rm M}_{\odot}$, 
$M_{\star} = 10^{11.13}~{\rm M}_{\odot}$, $\sigma =10^{2.308}$~km~s$^{-1}$,
$R_{\rm e}=4.86$~kpc and $n=4.74$. Assuming that the velocity dispersion profile
slope at $R_{\rm e}/2$ is $\eta=-0.053 \pm 0.044$, 
we try various anisotropies to see
how the resulting dark matter density profile depends on anisotropy.
We pick three cases of $\beta_{\rm mean}=+0.3,0,-0.3$ within $R_{\rm e}$.
For each value of $\beta_{\rm mean}$ we try ten different shapes of
the anisotropy, one constant case and nine other varying cases randomly selected
 from the ranges specified above that satisfy the Jeans equation and 
$\eta=-0.053$ up to the error of $0.044$. Fig.~\ref{btest} shows a total of 30
anisotropies and the corresponding dark matter profiles based on the
$\alpha$NFW model (eq.~4.6). First of all, comparing three constant cases we 
find that the resulting dark matter density profile varies systematically as a 
function of $\beta_{\rm mean}$;
$\alpha$ ($c_{\rm vir}$) decreases (increases) with $\beta_{\rm mean}$.
Secondly, for the case of $\beta_{\rm mean}=-0.3$ (colored blue) 
varying anisotropies have little effects on $\alpha$ and $c_{\rm vir}$. 
Thirdly, for zero mean anisotropy $\beta_{\rm mean}=0$ (colored red) varying 
shapes give on average significantly larger $c_{\rm vir}$
and a little lower $\alpha$ compared with the constant case. 
Finally, for the case of $\beta_{\rm mean}=+0.3$ (colored black) 
varying shapes can change $\alpha$ and $c_{\rm vir}$ significantly
(in some cases dramatically). Interestingly, the direction of movement in the
 parameter space due to varying anisotropy shapes from a constant is 
the same as that due to increasing $\beta_{\rm mean}$. 
This means that the effect of varying shape
can be mimicked by increasing $\beta_{\rm mean}$. 

As shown in the above case study a varying anisotropy can lead to a significant
change in the resulting dark matter density profile of an individual system.
Hence, if we kept the population mean of $\beta_{\rm mean}$ fixed, the 
statistical properties of resulting dark matter density profiles would
depend on whether we use constant or varying anisotropies. However, in our
procedure with constant anisotropy (section~4.3.2) we allowed $\beta_{\rm mean}$
to take values stochastically from a prior range, whereas
in the procedure with varying anisotropy used here 
we allowed shapes to vary stochastically within prior ranges 
while imposing a prior empirical distribution of $\beta_{\rm mean}$.
Because the effect of a varying anisotropy can be mimicked by 
changing $\beta_{\rm mean}$, it turns out that the statistical properties of dark
matter density profiles are similar between the two procedures with a (small) 
offset in $\beta_{\rm mean}$. This indicates that according to our stochastic 
procedure the resulting distribution of dark matter density profiles is
 minimally dependent on anisotropy models. Therefore, unless true anisotropy
shapes and/or the true mean of $\beta_{\rm mean}$ are significantly different from
those adopted here, it is unlikely that our assumption of constant 
anisotropy has significantly biased our results on dark matter density 
profiles.

\section{Dynamical mass scaling relations and alternative results for halo mass profile}

Our results on the halo profiles presented in the main text (section~4.3.2) 
are based on 
the observational constraints on the velocity dispersion profile (VP).
Dynamical mass scaling relations have been presented recently in the literature
\cite{Chu,Wol}. These relations are derived from the spherical 
Jeans equation in conjunction with physically well-motivated dynamical 
assumptions that are consistent with current observations.
 These relations allow us to estimate dynamical masses within certain radii
from velocity dispersions.
A crucial common feature of these relations is that they are 
insensitive to the value of the velocity dispersion anisotropy $\beta$.

The first mass estimate \cite{Chu} is expressed as 
\begin{equation}
M^{\rm (est)}(r_{\rm opt})=\frac{f_v^{-2}}{G}r_{\rm opt}\sigma_{\rm los}^2(r_{\rm opt}),
\end{equation}
where $r_{\rm opt}$ is an optimal radius defined in the reference and 
$r_{\rm opt} \sim 0.5 R_{\rm e}$ typically. The numerical factor $f_v$ is a 
velocity ratio defined in the reference and typically $f_v \sim 0.6$. This mass 
estimate is based on the assumption that the total gravitational potential is
that of the isothermal potential. This is not strictly satisfied by observed 
galaxies but a reasonable approximation for the potential within stellar
extents of galaxies supported by various astrophysical studies 
including strong lensing \cite{Rus,Koo,Bar}.

The second mass estimate \cite{Wol} is expressed as 
\begin{equation}
M^{\rm (est)}(r_{1/2})= \frac{3}{G}r_{1/2}\langle \sigma_{\rm los}^2 
\rangle(\infty), 
\end{equation}
where $r_{1/2}$ is the three-dimensional radius at which the enclosed stellar 
mass is $M_{\star}/2$ and $r_{1/2} \approx 4 R_{\rm e}/3$ for most stellar mass 
distributions of spheroids. This mass estimate is based on the assumption that 
$\sigma_{\rm los}(R)$ varies sufficiently slowly with $R$ near $R_{\rm e}$ as 
supported by current observations \cite{Cap2,Jor}.

\begin{figure}
\begin{center}
\setlength{\unitlength}{1cm}
\begin{picture}(15,11)(0,0)
\put(-0.7,12.){\includegraphics{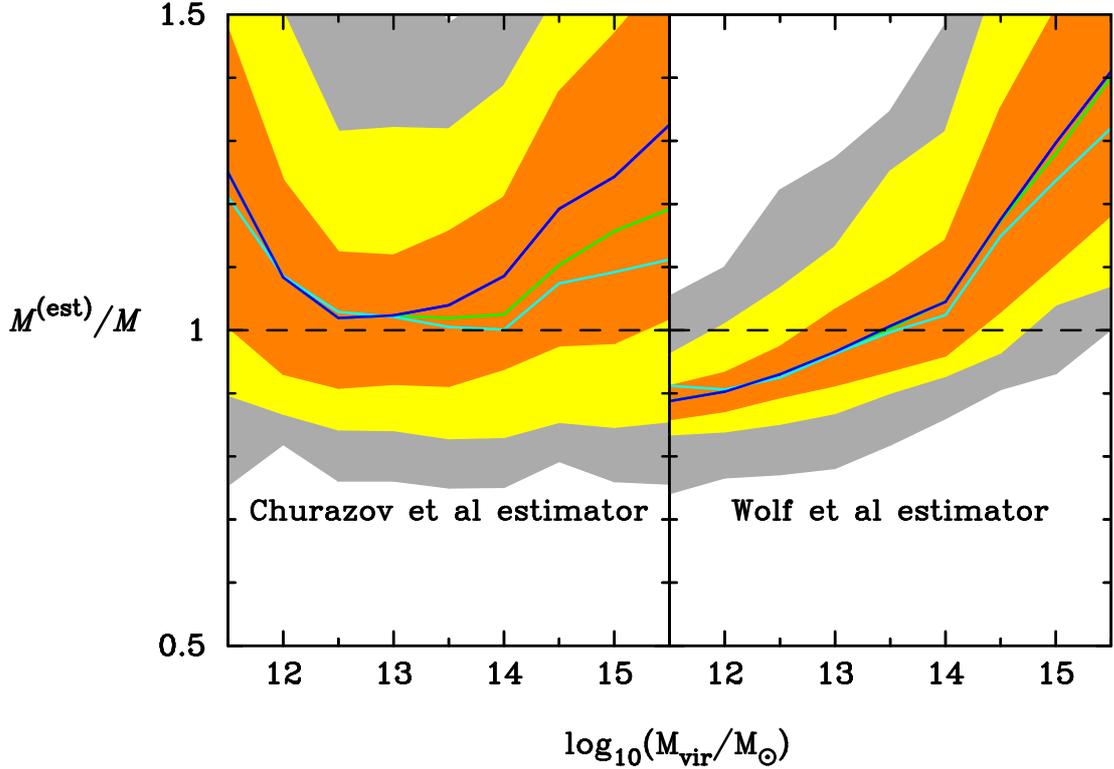}}
\end{picture} 
\caption{
Ratios of the estimated masses based on two scaling relations from the 
literature to the masses within two radii $r_{1/2}$ ($\approx 4 R_{\rm e}/3$, 
left panel) and 
$r_{\rm opt}$  ($\sim R_{\rm e}/2$, right panel) for our galaxy-halo systems
based on the observational VP constraints. The orange, yellow and gray
regions include the 68\%, 95\% and 99.7\% of our systems respectively.
The blue curves are the average values. The green and cyan curves are for
the corrected stellar masses as in Fig.~\ref{prof}.
}
\label{WCrat}
\end{center}
\end{figure}

Because of the approximate assumptions made we expect that our galaxy-halo
systems will satisfy the above mass scaling relations only approximately. 
Fig.~\ref{WCrat} shows the ratios of the estimated masses based on the scaling 
relations to the masses enclosed within the relevant radii for our systems. 
For $10^{12}~{\rm M}_{\odot}< M_{\rm vir}< 10^{14}~{\rm M}_{\odot}$ 
the mean estimated masses agree with the mean enclosed masses within 10\%. 
However, the mean ratio 
$\langle M^{\rm (est)}(r_{\rm opt})/M(r_{\rm opt}) \rangle$ 
(solid curves in the left panel of Fig.~\ref{WCrat}) tends to lie above unity
 whereas the mean ratio $\langle M^{\rm (est)}(r_{1/2})/M(r_{1/2}) \rangle$ 
(solid curves in the right panel of Fig.~\ref{WCrat})
systematically varies as a function of $M_{\rm vir}$. 
We do not speculate on possible sources of these behaviors.

\begin{figure}
\begin{center}
\setlength{\unitlength}{1cm}
\begin{picture}(15,11)(0,0)
\put(-0.7,12.){\includegraphics{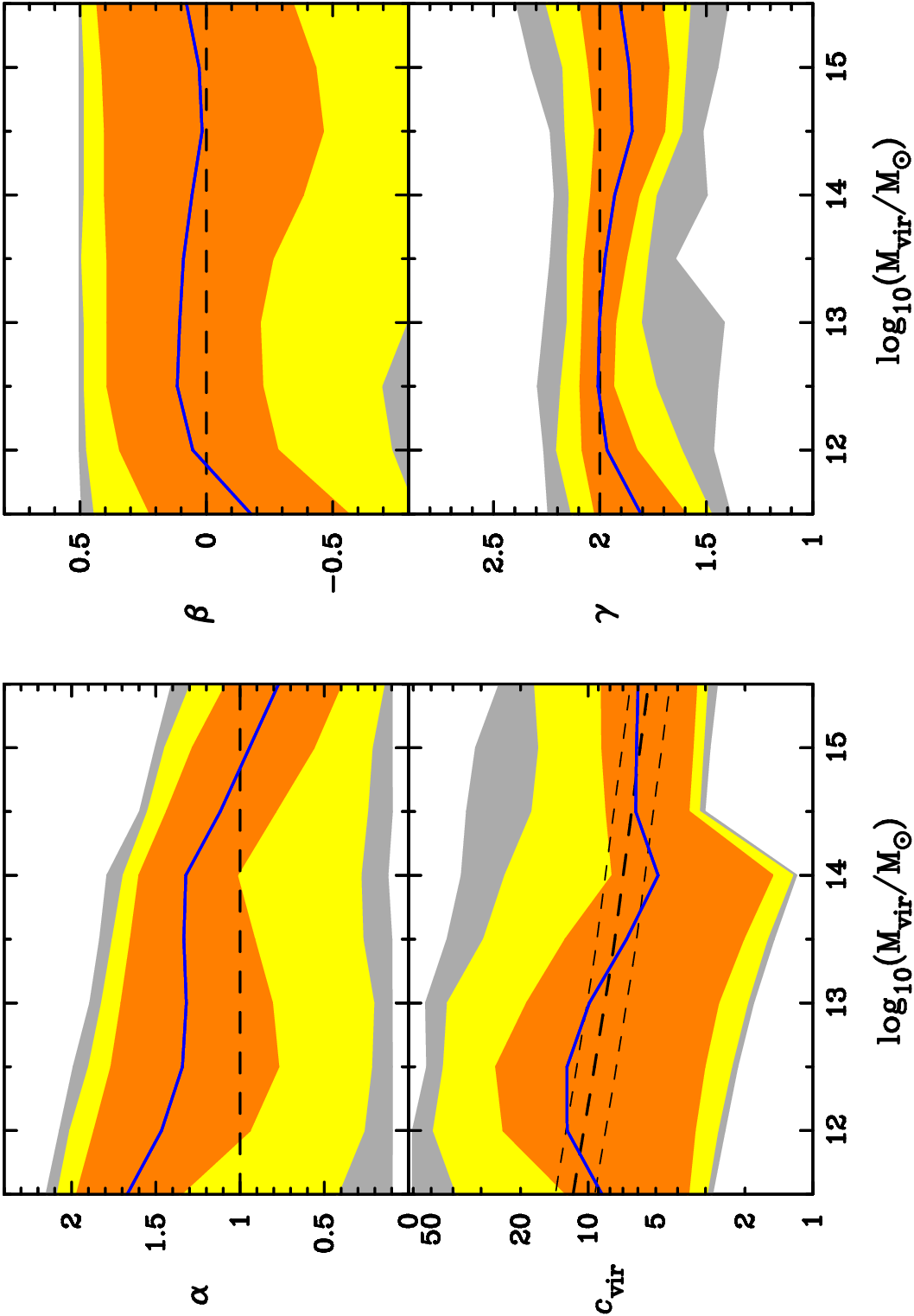}}
\end{picture} 
\caption{
Same as Fig.~\ref{prof} except that this result is based on
the combined figure-of-merit function of $Q^2_{\rm(VP)}+Q^2_{\rm(Chur)}$
[equation~(4.14) \& equation~(C.3)].
}
\label{profC}
\end{center}
\end{figure}

\begin{figure}
\begin{center}
\setlength{\unitlength}{1cm}
\begin{picture}(15,11)(0,0)
\put(-0.7,12.){\includegraphics{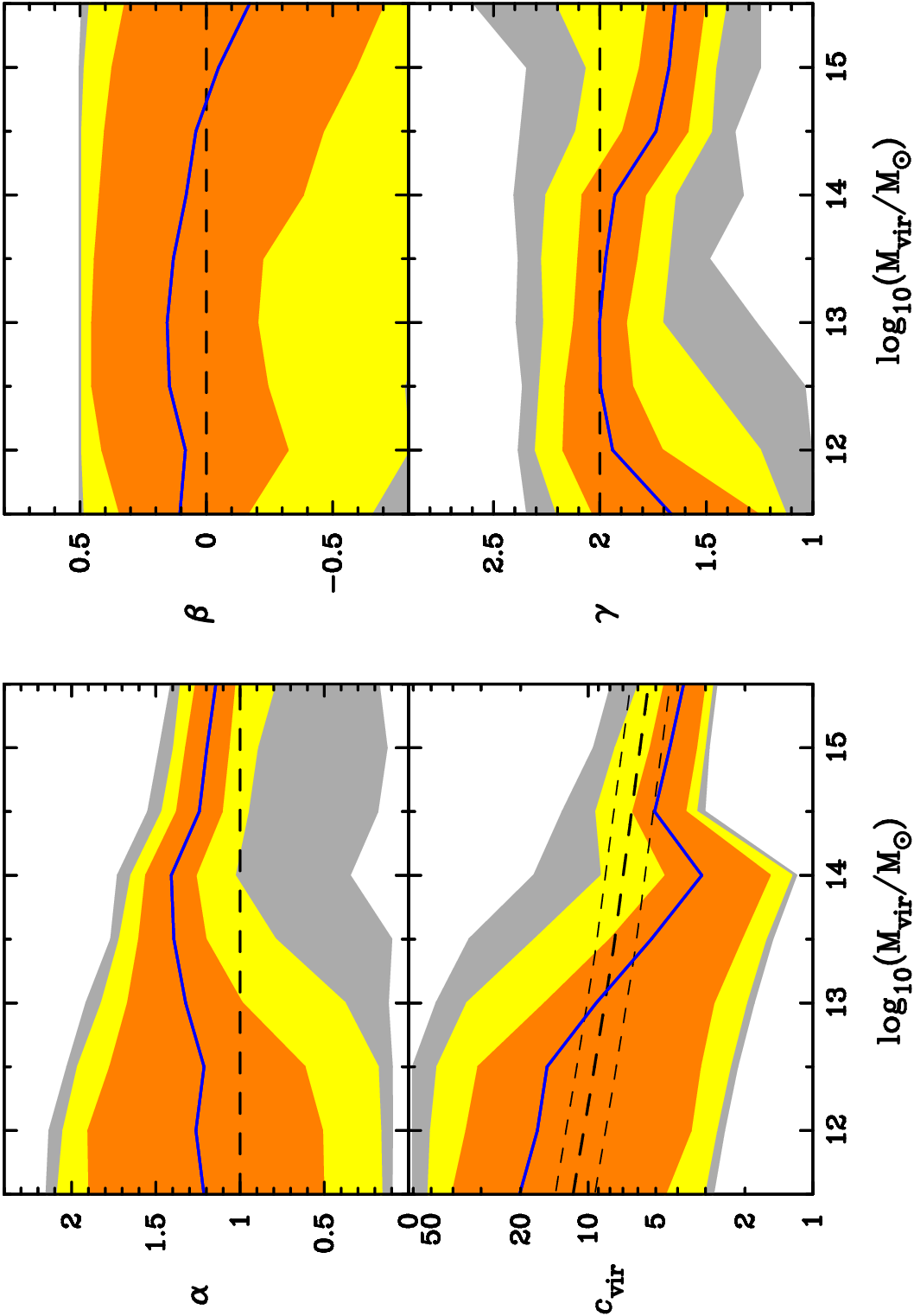}}
\end{picture} 
\caption{
Same as Fig.~\ref{prof} except that this result is based on
the combined figure-of-merit function of $Q^2_{\rm(VP)}+Q^2_{\rm(Wolf)}$
[equation~(4.14) \& equation~(C.4)].
}
\label{profW}
\end{center}
\end{figure}

In the above we have tested the dynamical mass scaling relations with our
systems constrained by the VP constraints only.
Alternatively, we could use the  dynamical mass scaling relations as
additional constraints on the halo dark matter density profiles. 
 We proceed by introducing two figure-of-merit functions defined by
\begin{equation}
Q^2_{\rm(Chur)} \equiv \frac{1}{\delta^2} 
\left(\frac{M^{\rm est}(r_{\rm opt})-M(r_{\rm opt})}{M(r_{\rm opt})}\right)^2, 
\end{equation}
and
\begin{equation}
Q^2_{\rm(Wolf)} \equiv \frac{1}{\delta^2} 
\left(\frac{M^{\rm est}(r_{1/2})-M(r_{1/2})}{M(r_{1/2})} \right)^2, 
\end{equation}
where we assume $\delta=0.1$.
We then combine these functions to the function $Q^2_{\rm(VP)}$  based on 
the VP constraints given by equation~(4.16) with $x$ replaced by 
$\langle\eta\rangle$. In doing so we perform a sort of least-square fitting
based on two independent constraints.
Fig.~\ref{profC} and Fig.~\ref{profW} show the results on the $\alpha$NFW model
 based on combined figure-of-merit functions given by 
$Q^2_{\rm(VP)}+Q^2_{\rm(Chur)}$ and $Q^2_{\rm(VP)}+Q^2_{\rm(Wolf)}$. 
Fig.~\ref{profC} is quite similar to Fig.~\ref{prof} except for 
$M_{\rm vir} \lesssim 10^{12}~{\rm M}_{\odot}$. 
Fig.~\ref{profW} shows some quantitative difference with Fig.~\ref{prof} 
in the behaviors of $\alpha$ and $c_{\rm vir}$.

\end{document}